\documentclass[12pt]{iopart}
\usepackage{graphicx}
\expandafter\let\csname equation*\endcsname\relax
\expandafter\let\csname endequation*\endcsname\relax
\usepackage{amsmath}
\usepackage{amssymb}
\usepackage{cite}
\usepackage{color,soul}
\usepackage{bm}% bold math

\input cyracc.def
\font\tencyr=wncyr10
\def\cyr{\tencyr\cyracc}

\begin{document}

\title[MICROSCOPE glitches]{MICROSCOPE mission: Statistics and impact of glitches on the test of the weak equivalence principle}

\author{Joel Berg\'e$^1$, Quentin Baghi$^2$, Alain Robert$^3$, Manuel Rodrigues$^1$, Bernard Foulon$^1$, Emilie Hardy$^1$, Gilles M\'etris$^4$, Sandrine Pires$^5$, Pierre Touboul$^1$}

\address{$^1$ DPHY, ONERA, Universit\'e Paris Saclay, F-92322 Ch\^atillon, France}
\address{$^2$ NASA Goddard Space Flight Center, Greenbelt, MD 20771, USA}
\address{$^3$ CNES, 18 avenue E Belin, F-31401 Toulouse, France} 
\address{$^4$ Universit\'e C\^ote d'Azur, Observatoire de la C\^ote d'Azur, CNRS, IRD, G\'eoazur, 250 avenue Albert Einstein, F-06560 Valbonne, France}
\address{$^5$ Laboratoire AIM, CEA, CNRS, Universit\'e Paris Saclay,  Universit\'e Paris Diderot, Sorbonne Paris Cit\'e, F-91191, Gif-sur-Yvette, France}
\ead{joel.berge@onera.fr}
\vspace{10pt}
\begin{indented}
\item[]December 2020
\end{indented}

\begin{abstract}
MICROSCOPE's space test of the weak equivalence principle (WEP) is based on the minute measurement of the difference of accelerations experienced by two test masses as they orbit the Earth. A detection of a violation of the WEP would appear at a well-known frequency $f_{\rm EP}$ depending on the satellite's orbital and spinning frequencies. Consequently, the experiment was optimised to miminise systematic errors at $f_{\rm EP}$. 
Glitches are short-lived events visible in the test masses' measured acceleration, most likely originating in cracks of the satellite's coating. In this paper, we characterise their shape and time distribution. Although intrinsically random, their time of arrival distribution is modulated by the orbital and spinning periods. They have an impact on the WEP test that must be quantified. However, the data available prevents us from unequivocally tackling this task. We show that glitches affect the test of the WEP, up to an {\it a priori} unknown level. Discarding the perturbed data is thus the best way to reduce their effect. 
\end{abstract}


\submitto{\CQG}
%\maketitle

\section{Introduction}

Hybridised with the MICROSCOPE satellite's star trackers \cite{robertcqg3}, the T-Sage's accelerometers \cite{liorzou20} are at the core of the satellite's drag-free and attitude control systems, allowing for the exquisite measurement and correction of minute non-gravitational accelerations, making MICROSCOPE one of the most gravitationally quiet laboratories in the Universe, within a small club of other space missions such as Gravity Probe B \cite{everitt11}, GOCE \cite{rummel11}, and LISA Pathfinder \cite{armano16, armano18}. Combined as two differential accelerometers, the two pairs of the T-Sage's accelerometers allowed for an unprecedented test of the Weak Equivalence Principle (WEP \cite{touboul17, touboul19}). 

However, since the drag-free system can only cancel low-frequency accelerations, transient accelerations (``glitches") can be observed in MICROSCOPE accelerometric data. Though generally of small amplitude (up to a dozen of nm/s$^2$ for individual T-Sage accelerometers), glitches are easily spotted and potentially contaminate the test of the WEP. Indeed, even if the test is performed with the difference of the accelerations measured by each sensor of a pair of accelerometers, small differences in their transfer functions mean that glitches do not perfectly cancel out.

Similar glitches were observed, albeit in much smaller numbers, in LISA Pathfinder data \cite{armano18, thorpe19}, but could not be satisfactorily explained yet.
Peterseim and colleagues \cite{flury08, peterseim12, peterseim_phd} reported the presence of many such glitches in the GRACE data \cite{tapley04}, some of them (dubbed ``twangs'') with no well-understood origin, but with some hints at correlations with the latitude (i.e., they occured when the GRACE satellites were about some prefered bands of latitude in an Earth geocentric frame). Nevertheless, the impact of twangs was shown to be small on the measurement of the geoid \cite{peterseim14}.

While the nature of LISA Pathfinder's and GRACE's glitches is already puzzling, we find the MICROSCOPE case even more difficult to grasp due to more complex modulation of the measurements with respect to the Earth (MICROSCOPE spinned about its axis normal to the orbital plane, while GRACE satellites were in an Earth-pointing configuration --always showing the same face to the Earth; LISA Pathfinder orbited the L1 Lagrange point, far from the Earth). 
In this paper, we do not try to pinpoint the exact physical processes behind glitches, but instead, we focus on their impact on the WEP measurement.

Beside glitches, some data points are missing in MICROSCOPE data, either lost from the telemetry or flagged as invalid by T-Sage's internal electronics, thus creating gaps in the data. Although short and infrequent, those gaps are prejudicious to the WEP-based data analysis \cite{bergecqg7}. 

In the absence of pre-flight reliable statistics on the occurence both of missing data and of glitches, we were not able to quantify {\it a priori} their impact on the test of the WEP, and we adopted a conservative approach where glitches are masked (thence, turned into missing data).
Techniques were then developed prior to MICROSCOPE's launch to allow for a precise and accurate least-square regression despite missing data \cite{baghi15, berge15, baghi16, pires16}. This strategy was actually our best bet.

Based on actual MICROSCOPE data, this paper aims to characterise the glitches' statistics and their impact on the test of the WEP.  In Sect. \ref{sect_anal}, we develop a linear model of the measurement apparatus and present a generic discussion of how short-lived events that share a similar shape and that follow a given time distribution can impact the test of the WEP. We characterise the time of occurence and the shape of glitches in Sect. \ref{sect_stats}, and discuss their impact on the power spectrum of the measured differential acceleration and on the test of the WEP in Sect. \ref{sect_WEP}. We conclude in Sect. \ref{sect_ccl}.

\section{Theoretical impact of glitches on the WEP measurement} \label{sect_anal}

\subsection{Differential acceleration}

\begin{figure}
\includegraphics[width=0.85\textwidth]{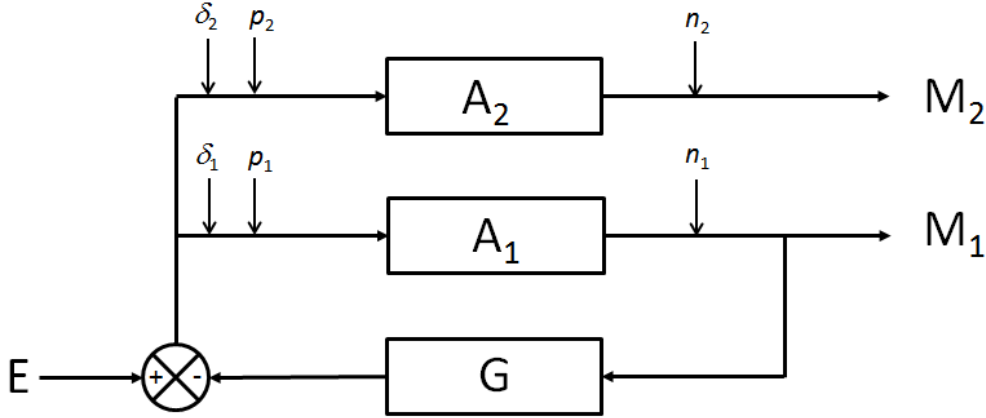}
\caption{Block diagram of the measurement process' linear model. The drag-free is controlled by $M_1$.}
\label{fig_block}       % Give a unique label
\end{figure}

The MICROSCOPE measurement relies on the intertwined control loops of both sensors and of the DFACS (Drag-Free and Attitude Control System). A block diagram of the system is shown in Fig. \ref{fig_block}, where the sensors' transfer functions are labeled $A_1$ and $A_2$, while $G$ is that of the DFACS. The drag-free is controled by only one sensor. The accelerations $M_1$ and $M_2$ measured by individual sensors are functions of the external accelerations $E$ (such as atmospheric drag, solar radiation pressure and glitches --we show in Sect. \ref{ssect_4masses} that glitches are events external from individual sensors), local accelerations ($\delta_i$ represent a composition-dependence of the ratio between the inertial and gravitational masses --$(m_g/m_i)_i=1+\delta_i$-- and $p_i$ other, generic perturbations), instrumental noise $n_i$ and of the closed-loop transfer functions. 

Assuming that the transfer functions of the two sensors have been matched up to an accuracy $dA$, such that $A_1(f)=A(f)$ and $A_2(f)= A(f)+dA(f)$, the acceleration measured by the sensor at the drag-free point is then
\begin{equation} \label{eq_acc1}
M_1 = \frac{A}{1+GA} (E + \delta_1 + p_1) + \frac{n_1}{1+GA}
\end{equation}
where all variables depend on the frequency, e.g. $G = G(f)$, and that measured at the other sensor is
\begin{equation} \label{eq_acc2}
M_2 = \frac{A+ dA}{1+GA}E + (A+dA)(\delta_2+p_2) - \frac{GA(A+dA)}{1+GA}(\delta_1 + p_1) - \frac{G(A+dA)}{1+GA}n_1 + n_2.
\end{equation}

The differential acceleration, from which we measure the WEP violation, is
\begin{multline} \label{eq_diffacc}
M_d \equiv M_1 - M_2 = -\frac{dA}{1+GA}E + A(\delta_1 - \delta_2) + A(p_1-p_2) +  \\
dA\left(\frac{GA}{1+GA}\delta_1 - \delta_2 \right) + dA \left(\frac{GA}{1+GA}p_1 - p_2\right) + n_1 - n_2 + \frac{GdA}{1+GA} n_1.
\end{multline}

If the transfer functions of both sensors are equal and perfectly matched ($dA = 0$), external forces do not impact the differential acceleration, thereby glitches have no effect on the estimation of the WEP violation.

The WEP is estimated at frequency $f_{\rm EP}$, where local accelerations $p_i \ll n_i$ (by design), and where, also by design $A(f_{\rm EP})\approx 1$, $G(f_{\rm EP})\approx 10^5$ \cite{robertcqg3}, and we assume that $dA(f_{\rm EP}) \ll A(f_{\rm EP})$. In this case, Eq. (\ref{eq_diffacc}) becomes
\begin{equation}
M_d(f_{\rm EP}) \approx -\frac{dA(f_{\rm EP})}{G(f_{\rm EP})A(f_{\rm EP})} E(f_{\rm EP}) + A(f_{\rm EP})\left[\delta_1(f_{\rm EP}) - \delta_2(f_{\rm EP}) \right] + n_1(f_{\rm EP}) - n_2(f_{\rm EP}). 
\end{equation}

The impact of glitches $\Delta$, is therefore simply bound by
\begin{equation} \label{eq_Delta}
|\Delta| = \left|\frac{dA(f_{\rm EP})}{G(f_{\rm EP})A(f_{\rm EP})} \pi(f_{\rm EP})\right| \leqslant \left| \frac{dA(f_{\rm EP})}{G(f_{\rm EP})A(f_{\rm EP})} E(f_{\rm EP})\right|
\end{equation}
where $\pi(f)$ is the frequency signature of glitches ($E = \pi + \,\, {\rm other\,\, external\,\, accelerations}$) and we assumed no cancellation between different external accelerations to provide the upper limit on the impact of glitches. 

Before investigating it in the following section, we can note that up to the noise, the acceleration measured at the drag-free point (Eq. \ref{eq_acc1}) provides $E(f_{\rm EP})/G(f_{\rm EP})$, when assuming $|\delta_1(f_{\rm EP})| \ll |E(f_{\rm EP})|$ and $|p_1(f_{\rm EP})| \ll |E(f_{\rm EP})|$. Then, an estimate of $dA(f_{\rm EP})/A(f_{\rm EP})$ is enough to provide an estimate of the impact of the total external accelerations.

\subsection{Frequency content of glitches: a generic discussion}

Eq. (\ref{eq_Delta}) shows that the impact of glitches on the test of the WEP is related to the power of glitches at the test frequency $f_{\rm EP}$. Before specialising to the MICROSCOPE case, we provide a rule of thumb analysis of what the frequency signature of glitches may look like.

We assume that the signal coming from glitches is some distribution of features external to the instrument (Sect. \ref{ssect_4masses}) and sharing a similar physical shape localized in time (the $m$th glitch starting at time $t_m$):
\begin{equation}
s_g(t) = \sum_m g_m(t)
\end{equation}

For simplicity, we assume that all glitches have the same shape $\chi$ but distinct amplitudes $a_m$, such that the $m$th glitch shape is observed in the measured time series as 
\begin{equation} \label{eq_glitch1}
g_m(t) = (a_m\chi * h)(t-t_m),
\end{equation}
where $*$ denotes the convolution product and $h(t)$ is the transfer function of the measuring apparatus (including the sensors and the DFACS).

The shape $\chi(t)$ is a priori unknown, as is the exact transfer function, preventing us from recovering $\chi$. However, we can rewrite Eq. (\ref{eq_glitch1}) as
\begin{equation} \label{eq_glitch2}
g_m(t) = (a_m\delta* k)(t-t_m),
\end{equation}
where $\delta(t-t_m)$ is the Dirac function and $k(t)$ is some kernel that provides the observed shape of glitches (albeit we do not have access to their real shape). We must emphasize that $k(t)$ is not the transfer function, but is a mathematical description that allows us to investigate the effect of glitches on the measurement and data analysis.

Recalling that the Power Spectral Density (PSD) of a stationary function $u$ is, according to Wiener-Khintchine theorem,
\begin{eqnarray}
S_u(f) & \equiv & {\mathcal F}\left\{ {\rm E}[u(t)u(t+\Delta t)] \right\}(f)  \\ %= |{\mathcal F}\left\{ s \right\}(f)|^2\\
& = & {\rm E}[|{\mathcal F}\{u\}(f)|^2],
\end{eqnarray}
where ${\mathcal F}\{u\}$ is $u$'s Fourier transform, and ${\rm E}[u(t)u(t+\Delta t)]$ is the signal's 2-point correlation function (${\rm E}$ denoting the expectation value).%, the PSD of the glitches (assumed to follow a stationary random process) is
Glitches are assumed to follow a stationary random process of order $N$, where $N$ is the number of points corresponding to some periodicity. Over one period, their averaged PSD is
\begin{equation} \label{eq_sg}
S_g(f) = {\mathcal F}\left\{ {\rm E}[\Xi(t)\Xi(t+\Delta t)] \right\}(f) \times |\tilde{k}(f)|^2,
\end{equation}
where $\Xi(t) = \sum_m a_m\delta(t-t_m)$, $\tilde{k}(f) \equiv {\mathcal F}\left\{ k \right\}(f)$ and we used both forms of the PSD to better highlight the two competing effects of glitches in the frequency domain: their time distribution and their shape.
Their time distribution (modulated by their amplitude) enters in the PSD as the Fourier transform of their 2-point correlation function while their shape is encoded in the Fourier transform of the kernel $k$, those two contributions being multiplied in the frequency domain. The kernel can therefore strongly impact the observed effect of glitches.

\begin{figure}
\center
\includegraphics[width=0.95\textwidth]{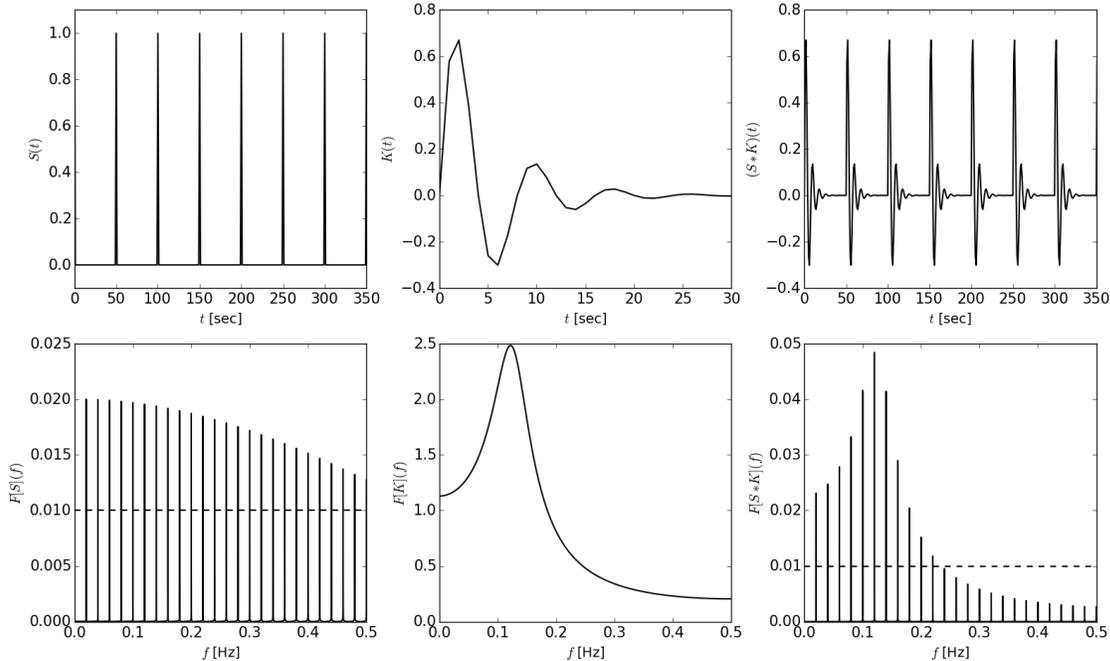}
\caption{Convolving a Dirac comb $S(t)$ with a damped sine $K(t)$ in the time domain (upper panel) is equivalent to multiplying their Fourier transforms in the frequency domain (lower panels). In the frequency domain, the convolution kernel dampens the power of the Dirac comb (due to the time distribution of impulses). The decreasing amplitude of the Dirac comb Fourier transform (lower left panel) is due to the finite observation window. The dashed line in the lower left and right panels shows a hypothetical noise (or detection) level.}
\label{fig_conv}       % Give a unique label
\end{figure}

As an illustration, we can assume that glitches come as a Dirac comb $ {\mbox{\cyr Sh}}_T(t)$ of period $T$ and that their observed shape is an exponentially damped sine $k(t) = \Theta(t)\sin(2\pi f_k t) \exp(-t/\tau)$, where $\Theta$ is Heaviside step function; as we show below, these assumptions are not far from reality.
In this case, the PSD of glitches is a Dirac comb of period $1/T$ multiplied by a Lorentzian (Fig. \ref{fig_conv}),
\begin{equation} \label{eq_sg2}
S_g(f) = \frac{\tau^2}{2\pi T^2} | {\mbox{\cyr Sh}}_{1/T}|^2  \left[ 4\pi^2 \tau^2 f_k^2 + (2 - 2\tau^2 f^2) (1+\tau^2 f^2)^2 \right]^{-1}.
\end{equation}
Note that the central frequency of the Lorentzian, its maximum value and its full width at half maximum depend on the kernel's parameters $f_k$ and $\tau$.

In this case, glitches can therefore impact the WEP analysis if
\begin{itemize}
\item the power of the Dirac comb is significant at the test frequency $f_{\rm EP}$ (e.g. if it has a period $1/f_{\rm EP}$).
\item and the glitches kernel does not bring the Dirac comb below the noise level at $f_{\rm EP}$.
\end{itemize} 
In the example of Fig. \ref{fig_conv}, if the dashed line (lower panels) shows the noise level, glitches would affect only the search for a signal of frequency $f \propto 0.02$Hz and lower than 0.22Hz.

In more realistic cases, glitches can come randomly in time, or as sums of approximate Dirac comb (i.e., affected by some jitter), each one with its own amplitude. Some combinations of different periods may create an impact at $f_{\rm EP}$, even if no Dirac with period $1/f_{\rm EP}$ exists. Nevertheless, we expect that the kernel of glitches originating from similar processes is constant; it may be that different kinds of glitches exist, each with its own kernel.
Such realistic cases should be treated numerically and taylored to the actually measured data, as done in the next section.

We shall summarise this section by emphasising that the effect of glitches on the measured PSD comes down to the multiplication of the Fourier transform of their time distribution by the Fourier transform of their observed shape. It is then important to be able to quantify their periodicities and model their shape.

\section{Statistics of MICROSCOPE glitches} \label{sect_stats}

In this section, we first describe the method used to detect glitches, before providing general statistics such as number distribution and density. We then investigate their geographical and time distributions, before showing their frequency content. The orbital period is always $T_{\rm orb}=5950$ seconds.

\begin{figure}
\includegraphics[width=0.55\textwidth]{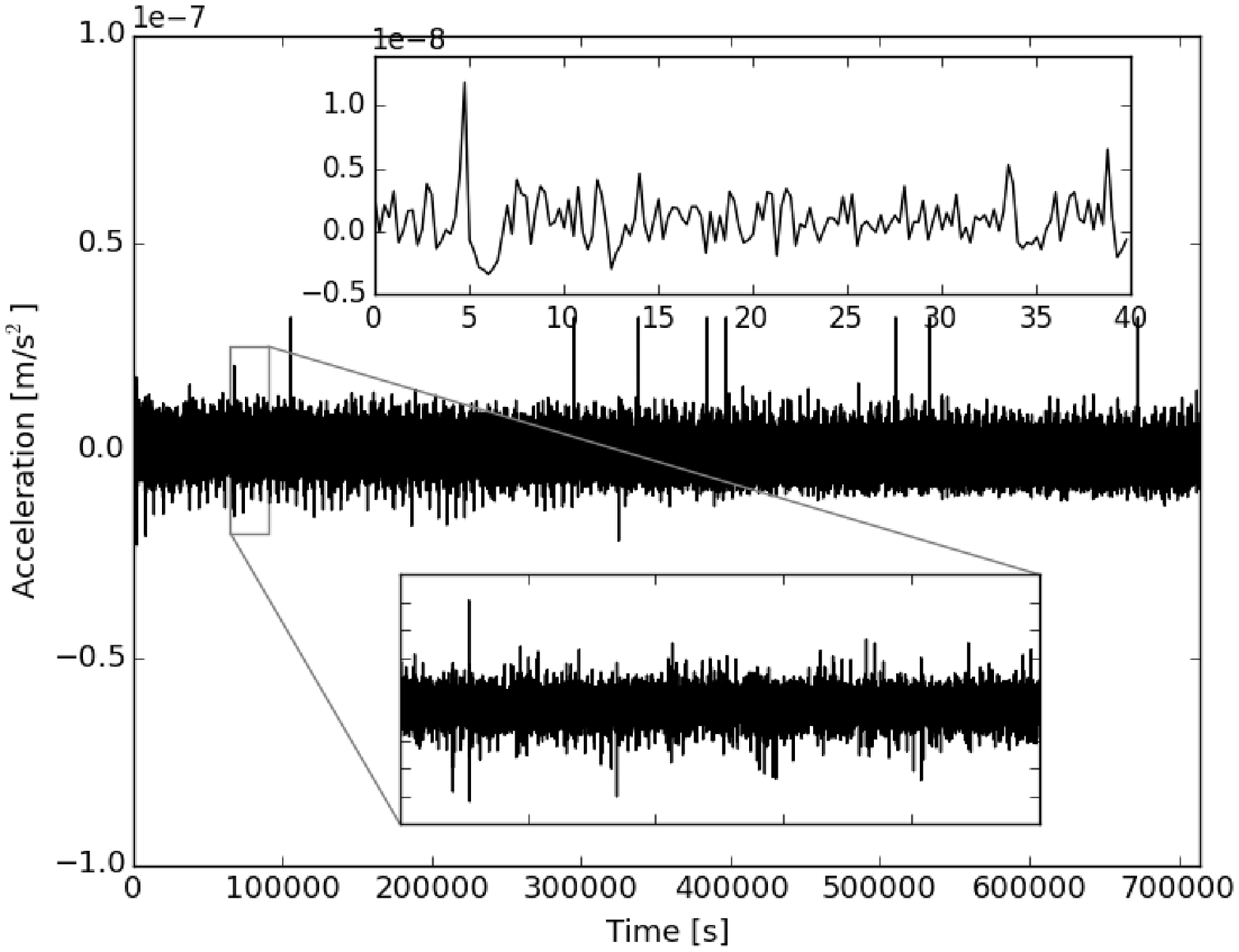}
\includegraphics[width=0.55\textwidth]{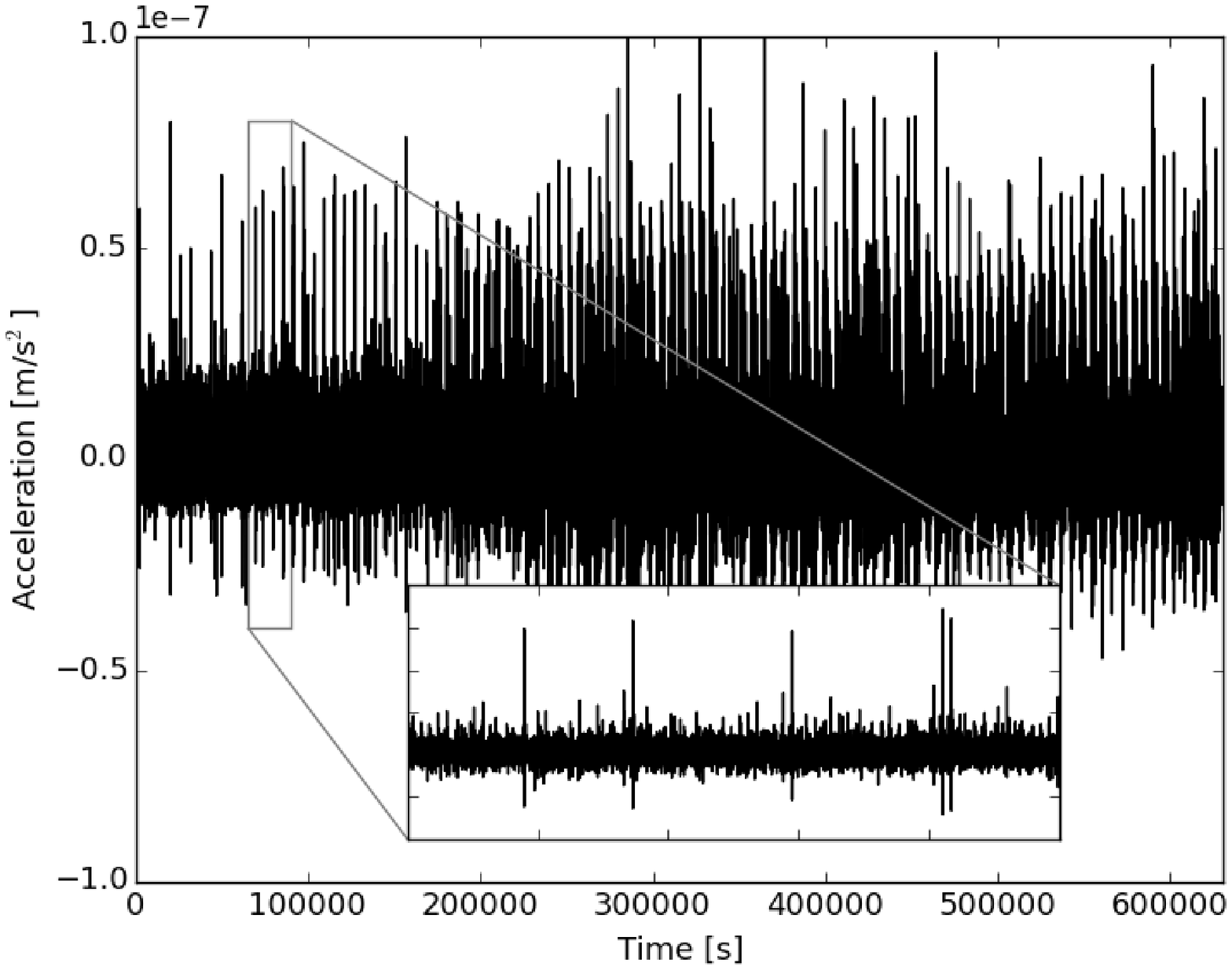}
\includegraphics[width=0.55\textwidth]{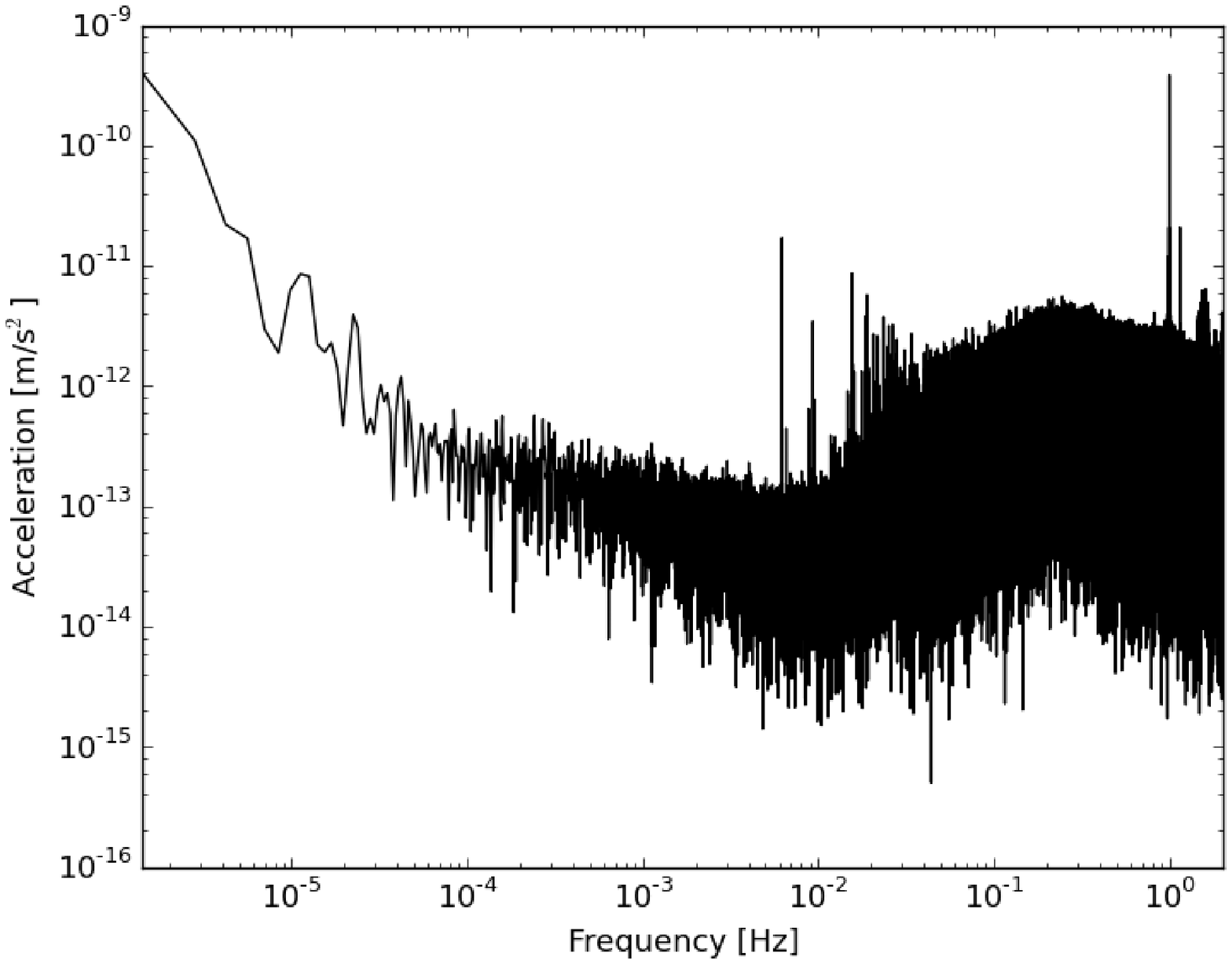}
\includegraphics[width=0.55\textwidth]{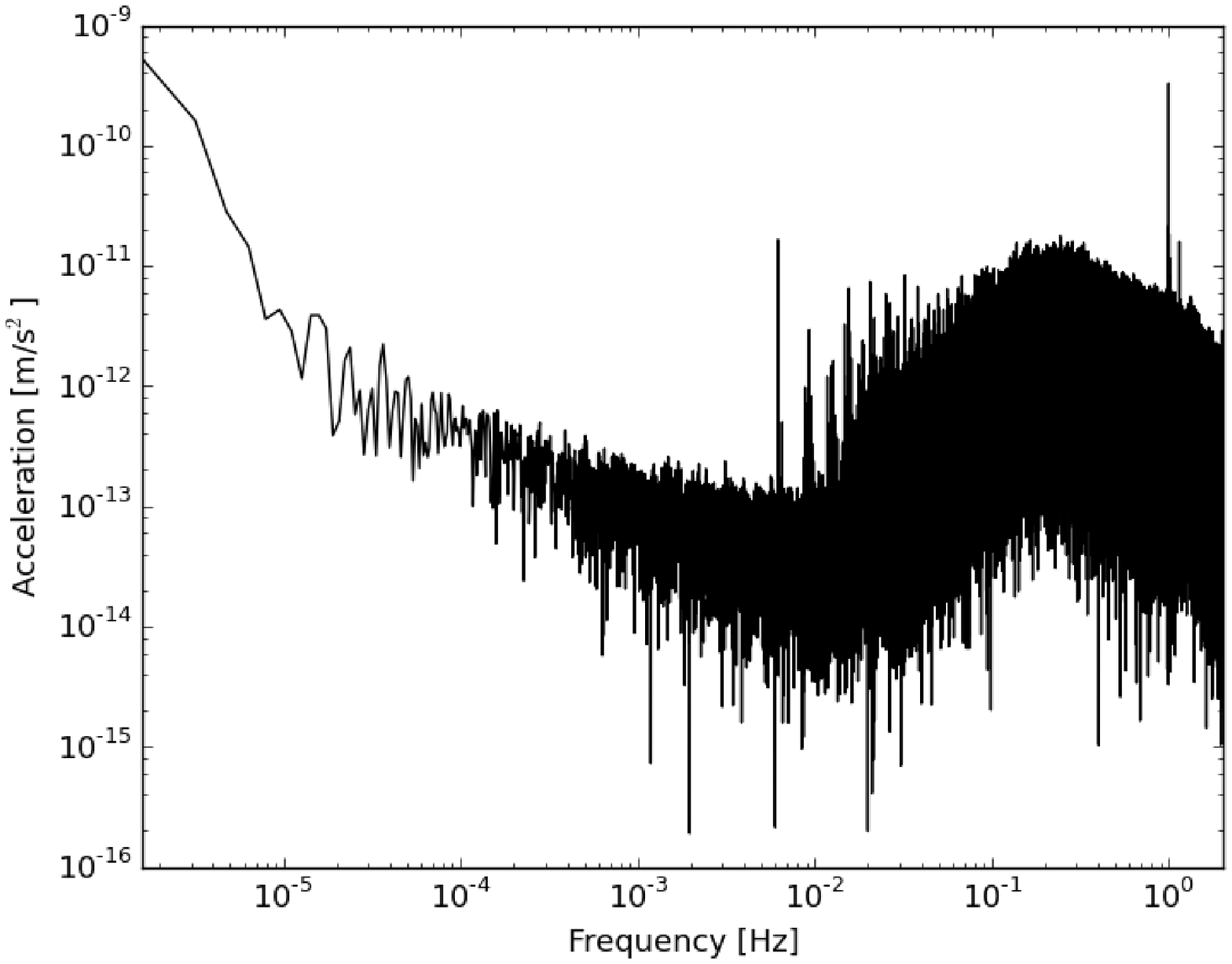}
\caption{Internal sensor's measured acceleration in the time (upper panels) and frequency (lower panels) domains for two sessions with quiet (left) and louder (right) high-frequency noise. Note that the acceleration ranges of the main plots are the same for both sessions. Lower insets show a zoom on 25000 seconds (approximately 5 orbits). The upper inset, restricted to 40 seconds, in the upper left panel shows a typical glitch; a smaller one is visible on the right.}
\label{fig_218252}       % Give a unique label
\end{figure}

Fig. \ref{fig_218252} shows the acceleration of the internal sensor of the SUEP instrument for two sessions, in the time and frequency domains. In those sessions, the satellite's spin rate $f_{\rm spin}=35/2 f_{\rm orb}$ which entails a rotation of the satellite on itself of period $T_{\rm spin} = 340$ seconds.
Zooms are provided in insets, which last 25000 seconds (approximately five orbits). A typical glitch is also shown in the upper left panel.
From this figure, it is clear that even under identical instrumental conditions (albeit the external experimental conditions may vary, e.g. due to the solar weather), the measured acceleration, as well as number and amplitude of glitches, can significantly vary (note that the acceleration range of the time domain is the same for both sessions). 

Missing and invalid data (unrelated to glitches) represent only a few data points per session, and their effect is circumvented with the techniques described in \cite{bergecqg7, baghi15, berge15, baghi16, pires16}. In the remainder of this paper, unless stated otherwise, we fill gaps and replace invalid data by a local average of the data. They are so rare that this technique does not bias the results.

\subsection{Glitches detection and general statistics}

Glitches are detected in two steps. First, we perform a recursive $\sigma$-clipping (e.g. Ref. \cite{akhlaghi15}) to extract outliers from the measured acceleration. Second, we segment the acceleration to gather outliers that belong to a single glitch (for instance, in the glitch shown in the inset of Fig. \ref{fig_218252}'s upper panel, the outliers at $t=4.5$s and $t=5.5$s clearly belong to the same outlier).

\subsection{Glitches are external events} \label{ssect_4masses}

\begin{figure}
\center
\includegraphics[width=0.75\textwidth]{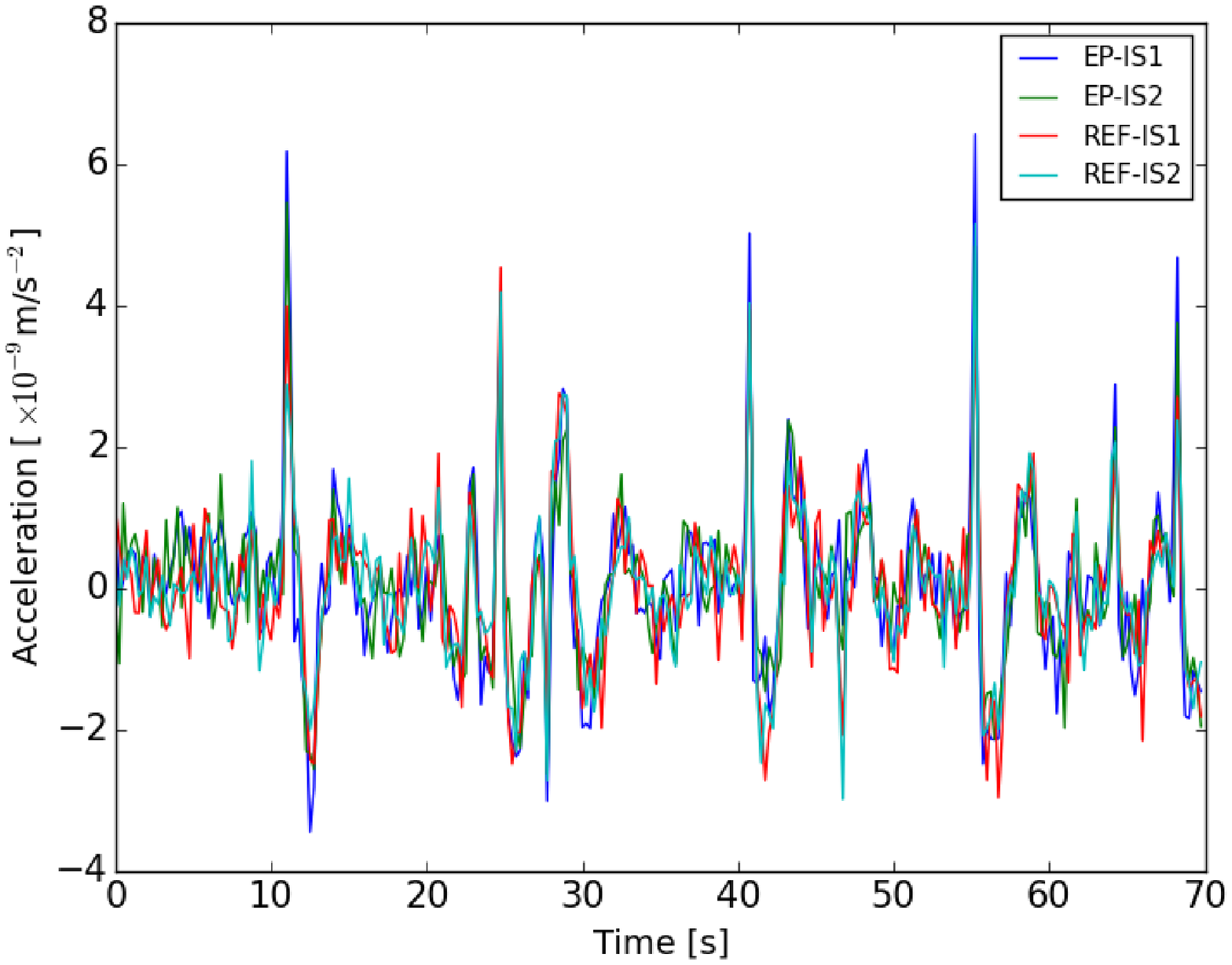}
\caption{Acceleration measured by all four test masses during a burst of glitches. Glitches are visible in the four accelerations.}
\label{fig_4masses}       % Give a unique label
\end{figure}

Glitches are events external to the instrument, since they appear simultaneously and with similar amplitudes on all four test-masses, as can be seen in Fig. \ref{fig_4masses}.

\subsection{Geographical and time distributions} \label{sect_statistics}

We now look for time and geographic periodicities in the occurrence of glitches. Evidently, such periodicities are linked since MICROSCOPE moves at a regular pace around the Earth; nevertheless, it is instructive to investigate them in parallel.

\subsubsection{Geographical distribution}

We start by investigating how glitches are distributed about the Earth. Note that we do not aim to investigate a potential systematic geographic distribution here. Some hints for a seasonal variation of the geographic distribution are given in \ref{app_geo}, where ``local'' effects (since each side of the satellite has its own propension to crack) are marginalised by averaging over sessions.
Instead, we aim to visualise all effects (local and geographical). Thus, we bin the surface of the Earth in latitude and longitude and compute the mean number of glitches per session in each bin with a grid of resolution $30\times 40$ pixels for individual sessions (to avoid marginalising local effects when averaging over different sessions). The high resolution is enough to see the effects of the satellite spinning as it orbits the Earth, even for the highest spin rate.

\begin{figure}
\center
\includegraphics[width=0.45\textwidth]{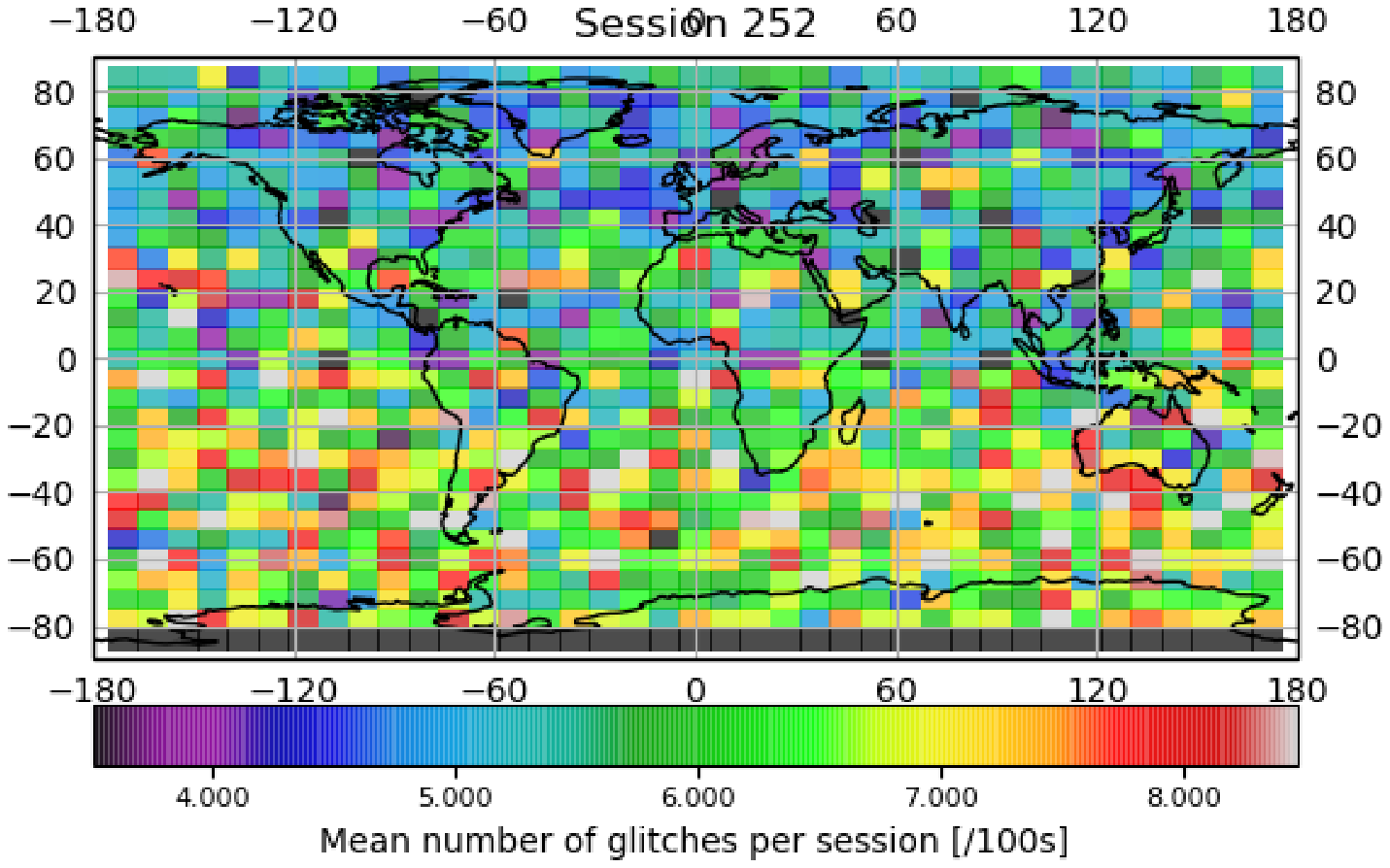}
\includegraphics[width=0.45\textwidth]{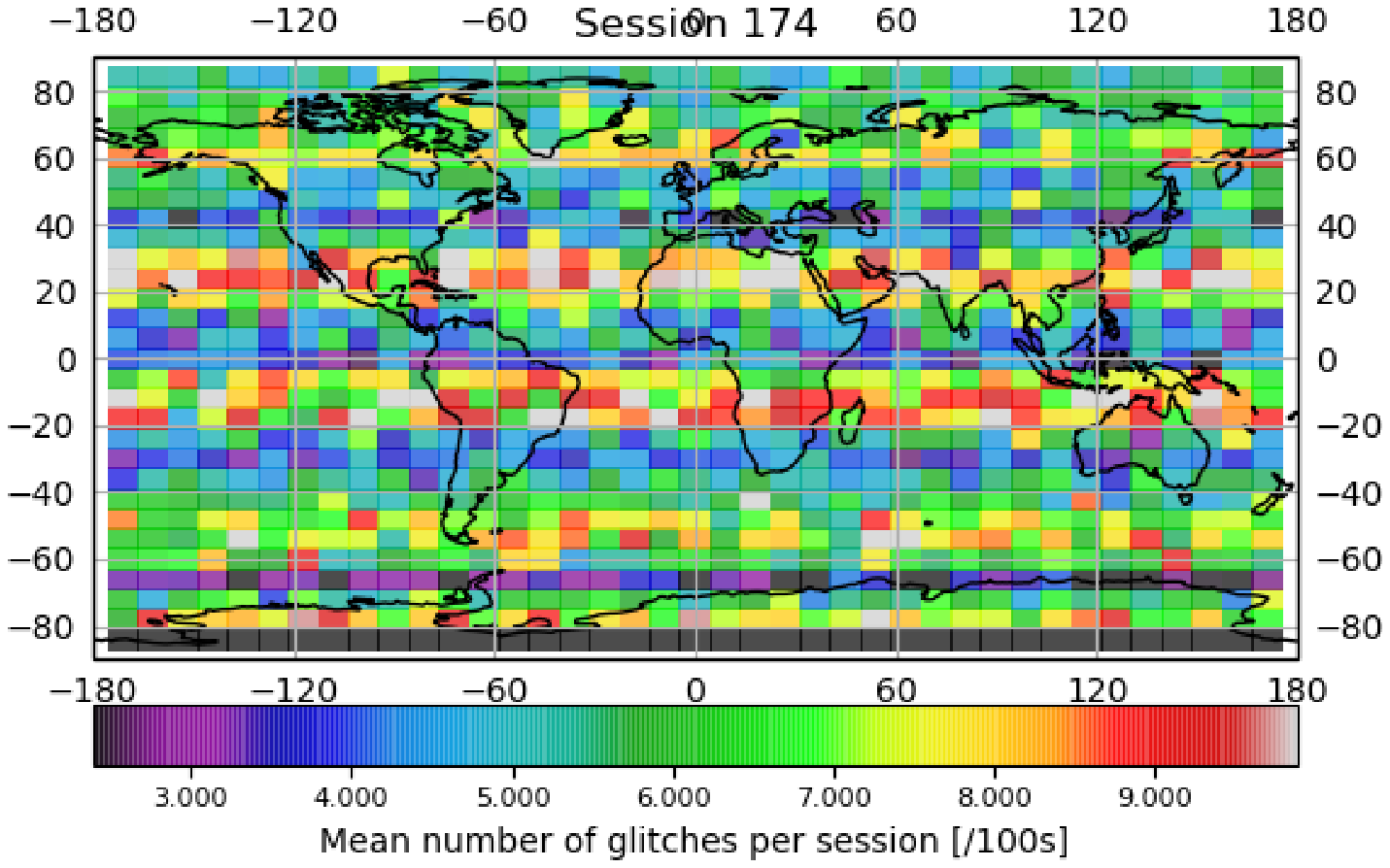}
\caption{Geographical distribution of glitches (per periods of 100 seconds) for two sessions with different spin rates (left: $f_{\rm spin}=35/2 f_{\rm orb}$; right: $f_{\rm spin}=9/2 f_{\rm orb}$).}
\label{fig_geo_ind}       % Give a unique label
\end{figure}

Fig. \ref{fig_geo_ind} shows the mean number of glitches for two sessions with different spin rates: $f_{\rm spin}=35/2 f_{\rm orb}$ ($T_{\rm spin}=340$s) for the left panel, and $f_{\rm spin}=9/2 f_{\rm orb}$ ($T_{\rm spin}=1322$s) for the right panel. A geographic dipole appears in the former (with glitches more likely to occur in the southern hemisphere), modulated by faint stripes corresponding to the distance traveled by the satellite during a rotation; the dipole is more difficult to see in the right panel, where stripes corresponding to the satellite's rotation are instead clearly visible. These geographical distributions hint to two periodicities: one linked to the orbital period and a second one related to the satellite spin rate. We checked that the same patterns appear in all accelerometers, hinting to an external cause for glitches.

\subsubsection{Time distribution}

Glitches follow an approximate Poisson process, albeit with a periodic clustering. Although this process is found to be stationary on the timescale of a given session, its mean rate evolves from session to session.
Computing the time distribution of glitches is similar to computing their clustering, which we quantify by means of the two-point correlation function $w(\Delta t)$: it gives the relative probability of finding a pair of glitches separated by a certain time interval $\Delta t$ with respect to that of a homogeneous Poisson distribution.
We use the estimator
\begin{equation}
w(\Delta t) = \frac{DD(\Delta t)}{RR(\Delta t)}-1,
\end{equation}
which is just a normalised version of the estimator used in Sect. \ref{sect_anal}. In this expression, $DD$ and $RR$ are the number of data-data and random-random pairs in a time bin $\Delta t$, where random glitches are drawn from a homogeneous Poisson law within the same time-span as the data. Were glitches distributed uniformally, $w(\Delta t)=0$; any excursion from $w(\Delta t)=0$ highlights a preferred scale.

We estimate the uncertainty on the 2-point correlation function $w(\Delta t)$ for a given session with the method presented in Ref. \cite{scranton02}: we divide the session into $N$ subsamples of equal length (long enough to include a significant number of glitches), on which we compute the correlation function $w_i(\Delta t)$. From those $N$ separate estimations, we calculate the mean $\bar{w}(\Delta t)$ and error on the mean $\Delta\bar{w}(\Delta t)$:
\begin{equation}
\left[\Delta\bar{w}(\Delta t)\right]^2 = \frac{1}{N} \sum_{i=1}^N \left[ \bar{w}(\Delta t) - w_i(\Delta t)\right]^2.
\end{equation}

\begin{figure}
\center
\includegraphics[width=0.45\textwidth]{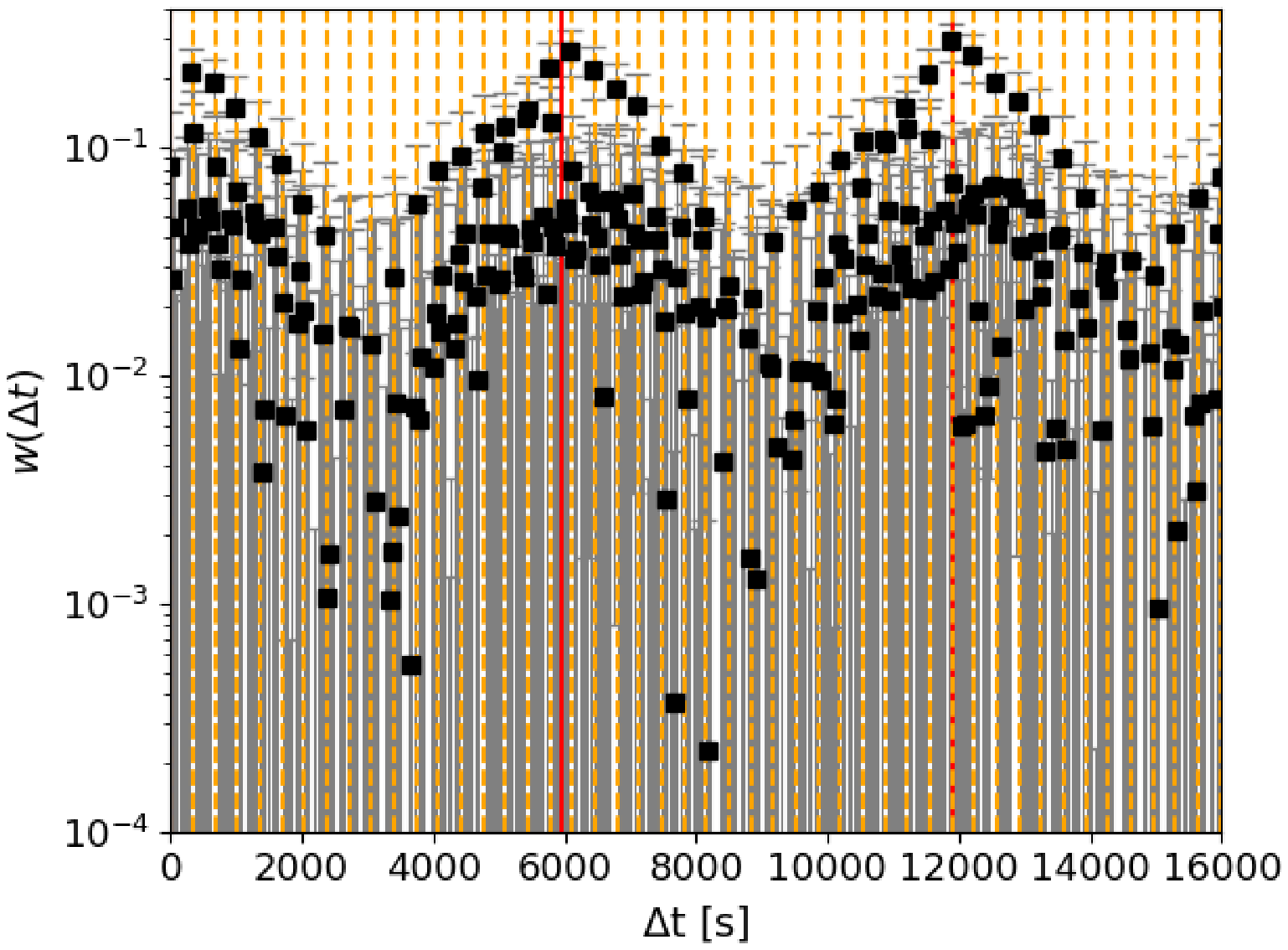}
\includegraphics[width=0.45\textwidth]{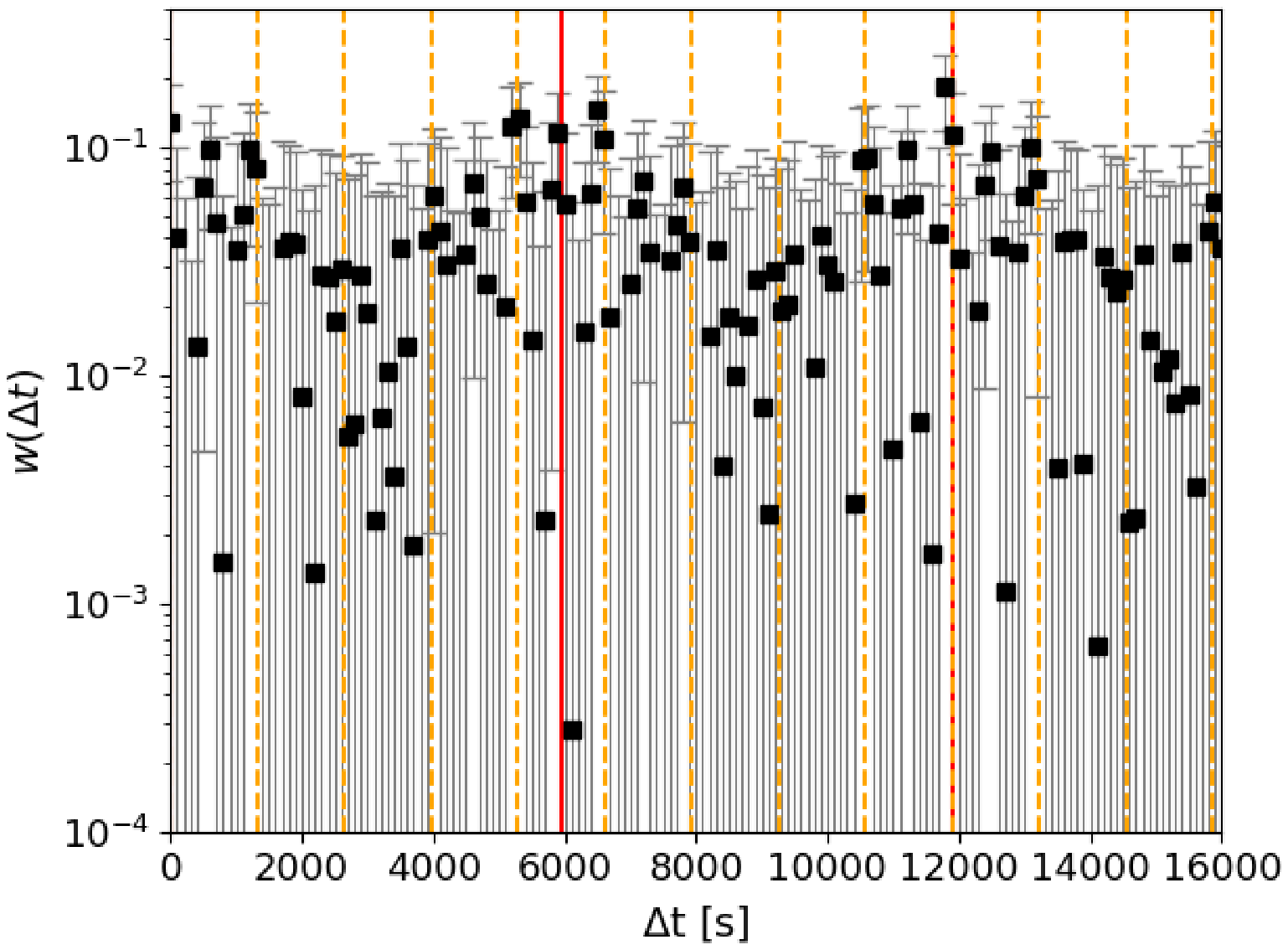}
\caption{Two-point correlation function of transients for the two sessions of Fig. \ref{fig_geo_ind}. The interval between consecutive vertical red solid lines shows the orbital period. The interval between consecutive vertical orange dashed lines shows the satellite's spin period.}
\label{fig_cf}       % Give a unique label
\end{figure}

Fig. \ref{fig_cf} shows the 2-point correlation function of glitches appearing in the SUEP internal sensor's acceleration for the sessions whose geographical distribution of glitches are shown in Fig. \ref{fig_geo_ind}.
For the sake of clarity, we restrict the range of the plot to a few orbit lengths; we checked that the same patterns is repeated all along the sessions durations. In this figure, the space between vertical red solid lines is equal to the orbital period $T_{\rm orb}$, while that between vertical orange dashed lines shows the satellite's spin period ($T_{\rm spin}=340$s  in the left panel, $T_{\rm spin}=1322$s in the right panel). Note that error bars (light grey) exclude 0 only close to time intervals corresponding to multiples of those periods: the correlation between other time intervals is negligible. The correlation function is computed in bins small enough to accommodate 10 points between consecutive vertical dotted line.

Two modulations are visible in both sessions: one at (approximately) the orbital period, and a second one at the spin period. They are intertwined, in the sense that the former is rather at twice the orbital period. This comes from the fact that the spin rate is a half-integer of the orbital frequency, meaning that the satellite-Earth system comes back in the same configuration (attitude vs latitude) every other revolution. 

These observations hint towards correlations between the occurence of glitches with both the satellite's position on its orbit and its orientation with respect to the Earth. They concur with the conclusions drawn from the geographical distribution, and support the hypothesis that glitches originate from crackles of the multi-layor insulator (MLI) coating of the satellite, with one side more sensitive than others. The exact source of excitation of the MLI is not well understood, but may be related to the Earth albedo or to the Sun illumination. The hypothesis that crackles of the MLI cause glitches is also supported by our experience with other missions: similar transients affected GRACE but GOCE data was exempt from them; GRACE had a MLI coating but GOCE did not. Most importantly, crackles were identified in several types of MLI during pre-launch test, and the MLI with least crackles was selected \cite{robertcqg3}.

\subsection{Glitches frequency content}

We saw in Sect. \ref{sect_anal} (Eq. \ref{eq_sg}) that both the time distribution and the shape of glitches affect their frequency content. The former as the Fourier transform of their 2-point correlation function, and the latter as the signature of the kernel transforming an impulse into an observed glitch (recall that this kernel is related to the instrument's transfer function, though does not represent it). We discuss those two contributions here.

\subsubsection{Frequency distribution}

\begin{figure}
\center
\includegraphics[width=0.75\textwidth]{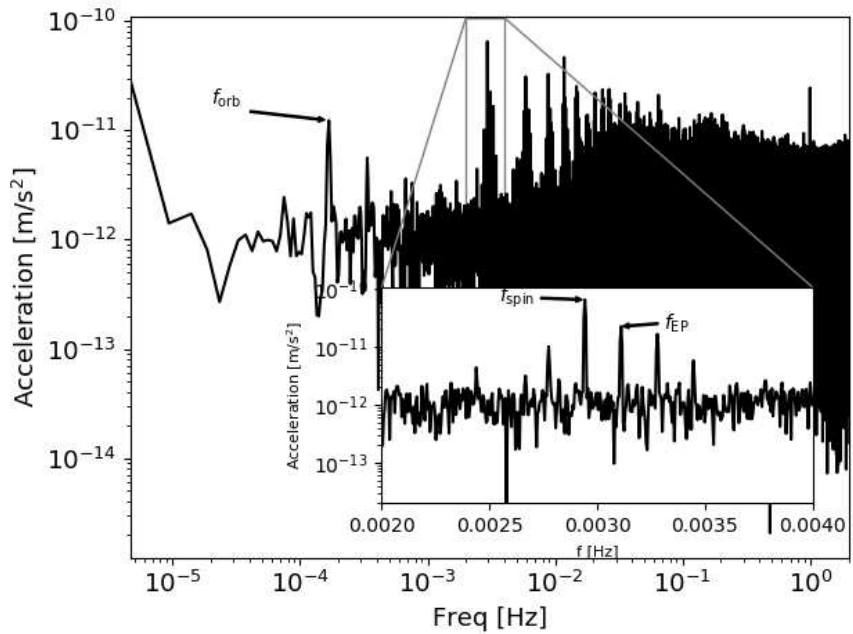}
\caption{FFT of the time distribution of SNR$>3$ glitches for a typical measurement session.}
\label{fig_diracfft}       % Give a unique label
\end{figure}

We first compute the Fourier transform of Eq. (\ref{eq_sg})'s $\Xi$ function. For a given measurement session, we extract glitches from the acceleration measured by one sensor, retrieve their amplitude, and define $\Xi(t)$ as the time-series consisting of Dirac impulses at the positions of glitches, with their corresponding amplitude.
Fig. \ref{fig_diracfft} shows the Fourier transform of the 2-point correlation of $\Xi$. The inset provides a zoom about the spinning and WEP test frequencies $f_{\rm spin}$ and $f_{\rm EP}$.

Consistently with our time-domain analysis above, we notice that glitches provide power at the orbital and spinning frequencies. The combination of those two periods brings power to other frequencies, distributed around $f_{\rm spin}$ and its harmonics, creating a forest of groups of spectral lines, mainly visible between a few mHz and 0.1 Hz. Noticeably, the WEP frequency is affected, with significant power brought up by the time distribution of glitches.

\subsubsection{Kernel shape} \label{ssect_shape}

The kernel $k$ introduced in Eq. (\ref{eq_sg}) represents glitches. Thus, they allow us to estimate it, though on a limited frequency range limited by the length of glitches (i.e., the observed shape gives no information on the behaviour of the kernel at frequencies smaller than $5\times10^{-2}$Hz).

\begin{figure}
\center
\includegraphics[width=0.45\textwidth]{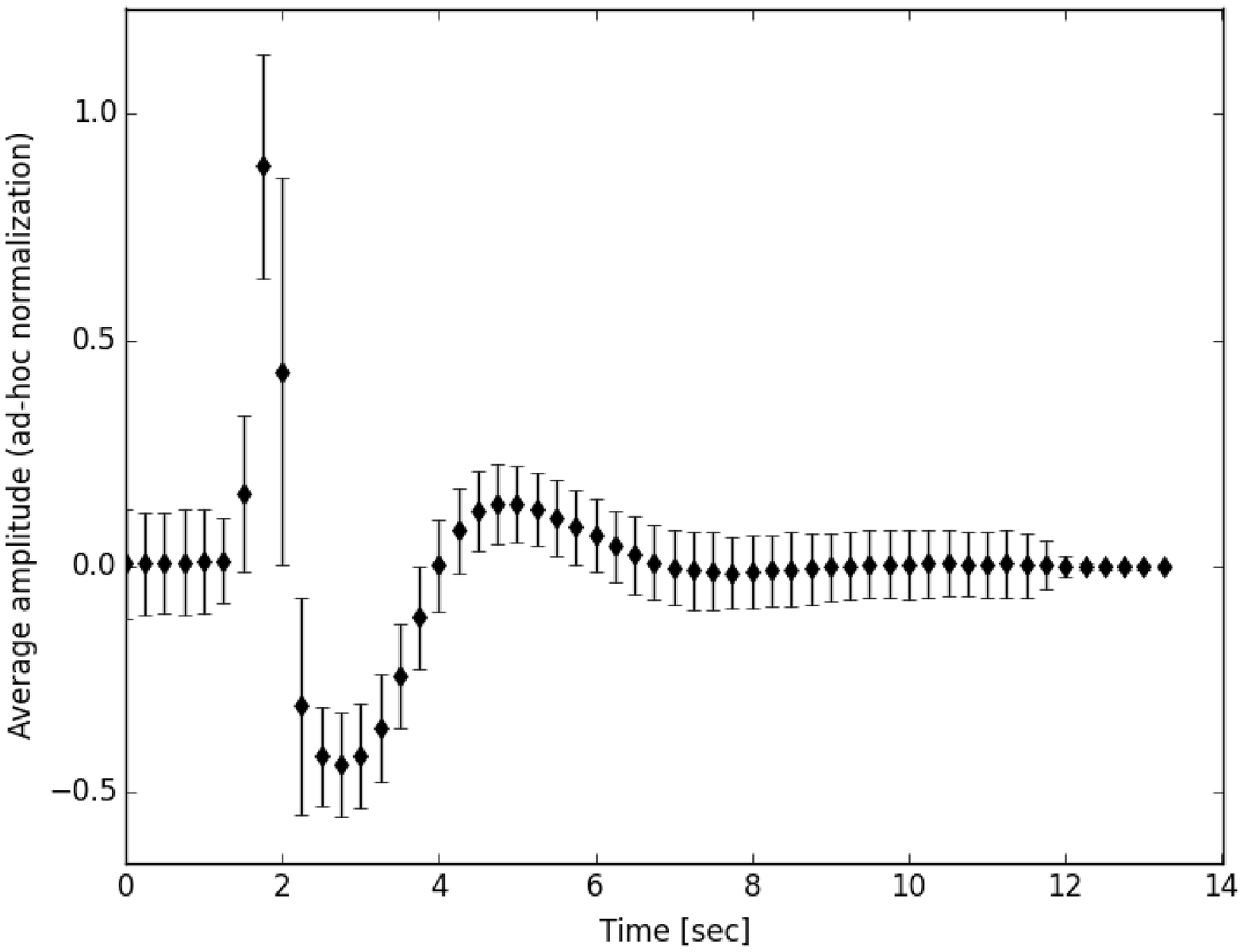}
\includegraphics[width=0.45\textwidth]{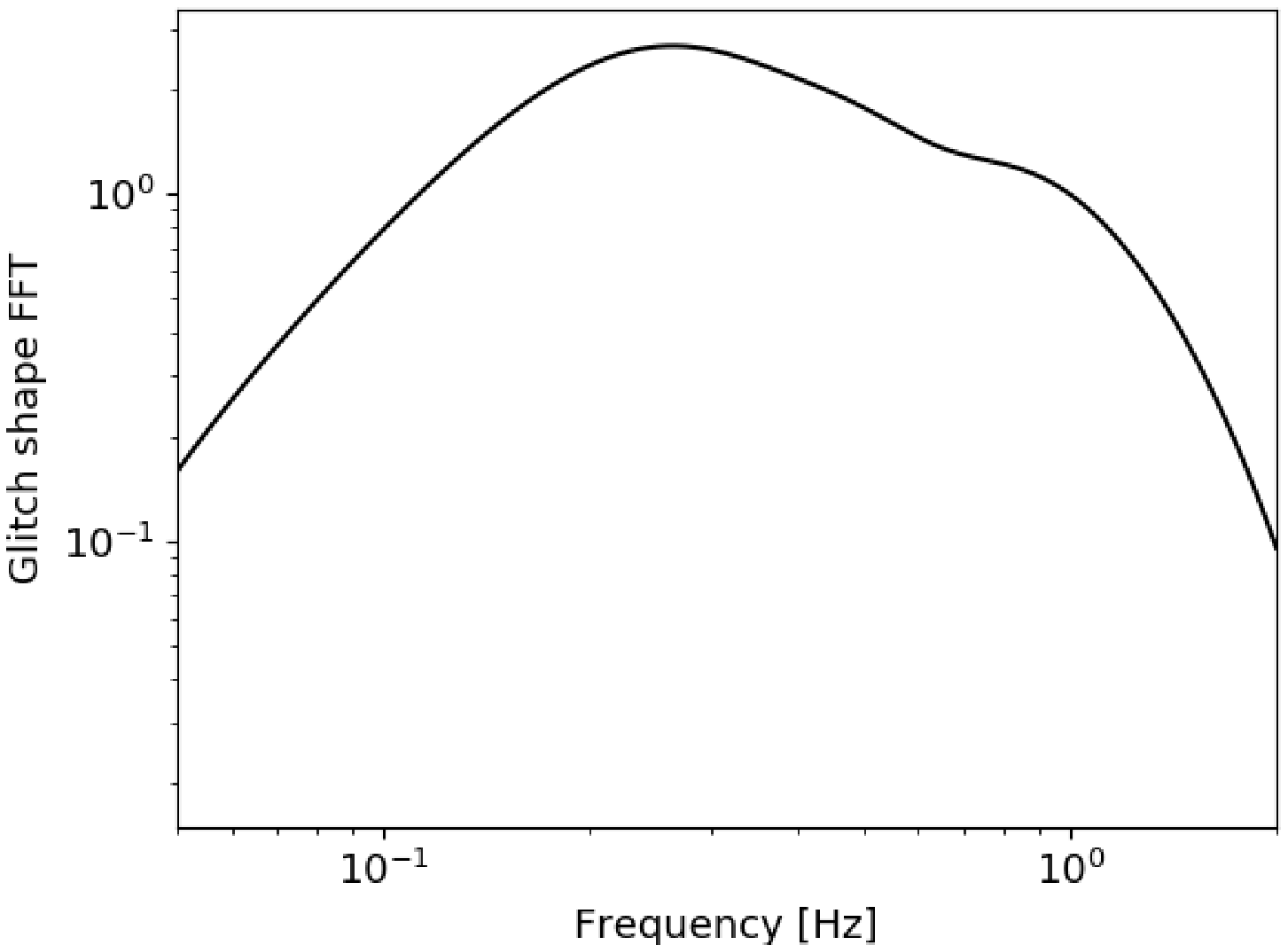}
\caption{Average acceleration glitch shape by sensor, with an ad-hoc normalisation. Left: time domain. Right: frequency domain.}
\label{fig_shape}       % Give a unique label
\end{figure}

We performed a Principal Component Analysis (e.g. Ref. \cite{hastie01}) on all glitches of Signal-to-Noise Ratio ${\rm SNR}>3$, for several measurement sessions. Only two components were found to be significant, each having the sign opposite to the other's. Therefore, all glitches have the same shape, multiplied by their relative amplitude.
The left panel of Fig. \ref{fig_shape} shows the average shape of glitches in the time domain. After a sudden increase, it oscillates but quickly decreases, recalling an exponentially damped sinusoid. The error bars correspond to the error on the mean.

The right panel of Fig. \ref{fig_shape} shows the FFT of this kernel. Although averaging glitches removes most of the noise, residual noise affects the FFT, as does the limited number of data points. To bypass those limitations, we fit the average glitch with exponential shapelets \cite{berge19}: this procedure allows us to denoise the average glitch and resample it at will, and to obtain the smooth Fourier transform shown in Fig. \ref{fig_shape}.
We can note that this kernel, though peaking about a central frequency, has a more complex shape than a Lorentzian. Glitches are thus more complicated than pure damped sines (Eq. \ref{eq_sg2}).

\subsubsection{Full frequency content} \label{ssect_ffc}

As shown by Eq. (\ref{eq_sg}), the frequency signature of the glitches is given by the product of the signature of their distribution with the kernel investigated above. We should emphasise that this kernel contains the contribution of the electrostatic sensor and of the drag-free system and covers the entire frequency range from $[10^{-5}-2]$ Hz. Thus, the shape shown above by no means represents its integrality. However, no in-flight experiment can give us access to the kernel's behaviour at frequencies lower than $5\times10^{-2}$Hz, preventing us from directly quantifying the effect of glitches over the entire frequency range. Thence, we must resort to a numerical model of the transfer function of the instrument.

From Eqs. (\ref{eq_glitch1}) and (\ref{eq_glitch2}), the (normalised and denoised) observed shape of glitches is given by $\chi_{\rm obs} = \chi_{\rm true}*h = \delta * k$, where we added the subscript ``true'' to emphasise that this function is the physical shape of the glitch (though sampled at 4 Hz). Noting that $\mathcal{F}\{\delta\}=1$, we can estimate the ``true'' shape of glitches as $\hat{\chi}_{\rm true}(t) = \mathcal{F}^{-1}\left\{\tilde{k}(f)/\tilde{h}(f)\right\}(t)$,
where $\tilde{k}$ and $\tilde{h}$ are the Fourier transforms of the kernel $k$ and of the instrument's transfer function $h$, restricted to frequencies available from the glitches shape ($0.05{\rm Hz} \leqslant f \leqslant 2{\rm Hz}$). This frequency restriction is prejudicial, and prevents us from recovering the correct ``true'' shape of glitches.

Although most glitches presumably originate from crackles of the satellite's MLI, some of them also come from crackles (called clanks in Ref. \cite{robertcqg3}) of the gas tanks as their pressure decreases while the gas is consumed. Pre-launch tests showed that those clanks are extremely short events, lasting about a few dozens milliseconds \cite{robertcqg3}. Since they cannot be singled out in MICROSCOPE data, we conclude that all glitches are extremely short events, much shorter than the response time of MICROSCOPE's electronics. Thus, they can safely be considered as Dirac impulses when seen in MICROSCOPE's 4Hz data. In the remainder of this paper, we assume that their true shape is a Dirac, so that we can compute the effect of glitches at all frequencies, from Eq. (\ref{eq_sg}), as
\begin{equation} \label{eq_sg3}
S_g(f) = {\mathcal F}\left\{ {\rm E}[\Xi(t)\Xi(t+\Delta t)] \right\}(f) |\tilde{h}(f)|^2,
\end{equation}
where we also make use of the transfer function model (valid at all frequencies) and do not rely on the empirical kernel anymore.

\begin{figure}
\center
\includegraphics[width=0.75\textwidth]{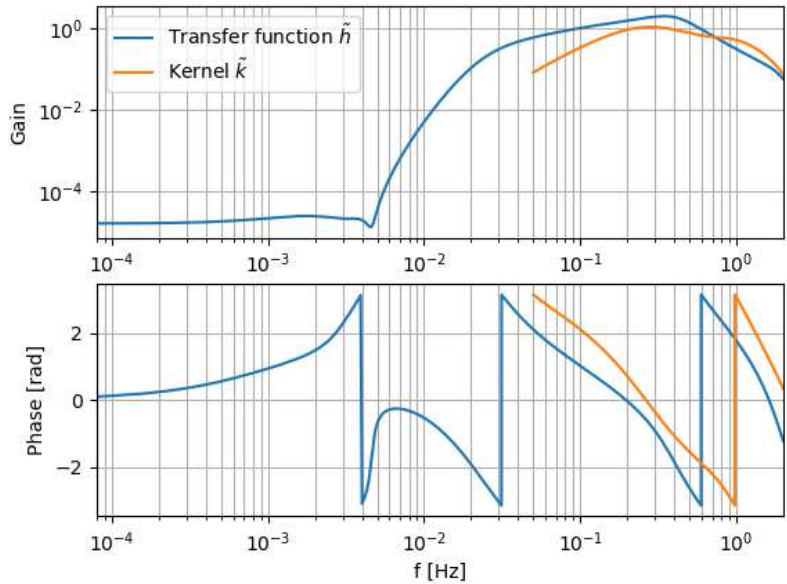}
\caption{Bode diagram of the model of the transfer function $h$ and of the kernel $k$ measured from the glitches shape.}
\label{fig_tf}       % Give a unique label
\end{figure}

As shown in Sect. \ref{sect_anal}, in the linear model considered thus far, the transfer function consists of the contribution of the electrostatic sensor and of the DFACS, such that the transfer function of the mass controlling the drag-free is
\begin{equation}
\tilde{h}(f) = \frac{A(f)}{1+G(f)A(f)},
\end{equation}
where we take the drag-free and sensor transfer functions from Refs. \cite{robertcqg3, chhuncqg5}.
The transfer function is shown in Fig. \ref{fig_tf}, and shows a good agreement with the kernel $\tilde{k}$ restricted to the frequencies available from the limited size of glitches. 

\begin{figure}
\center
\includegraphics[width=0.45\textwidth]{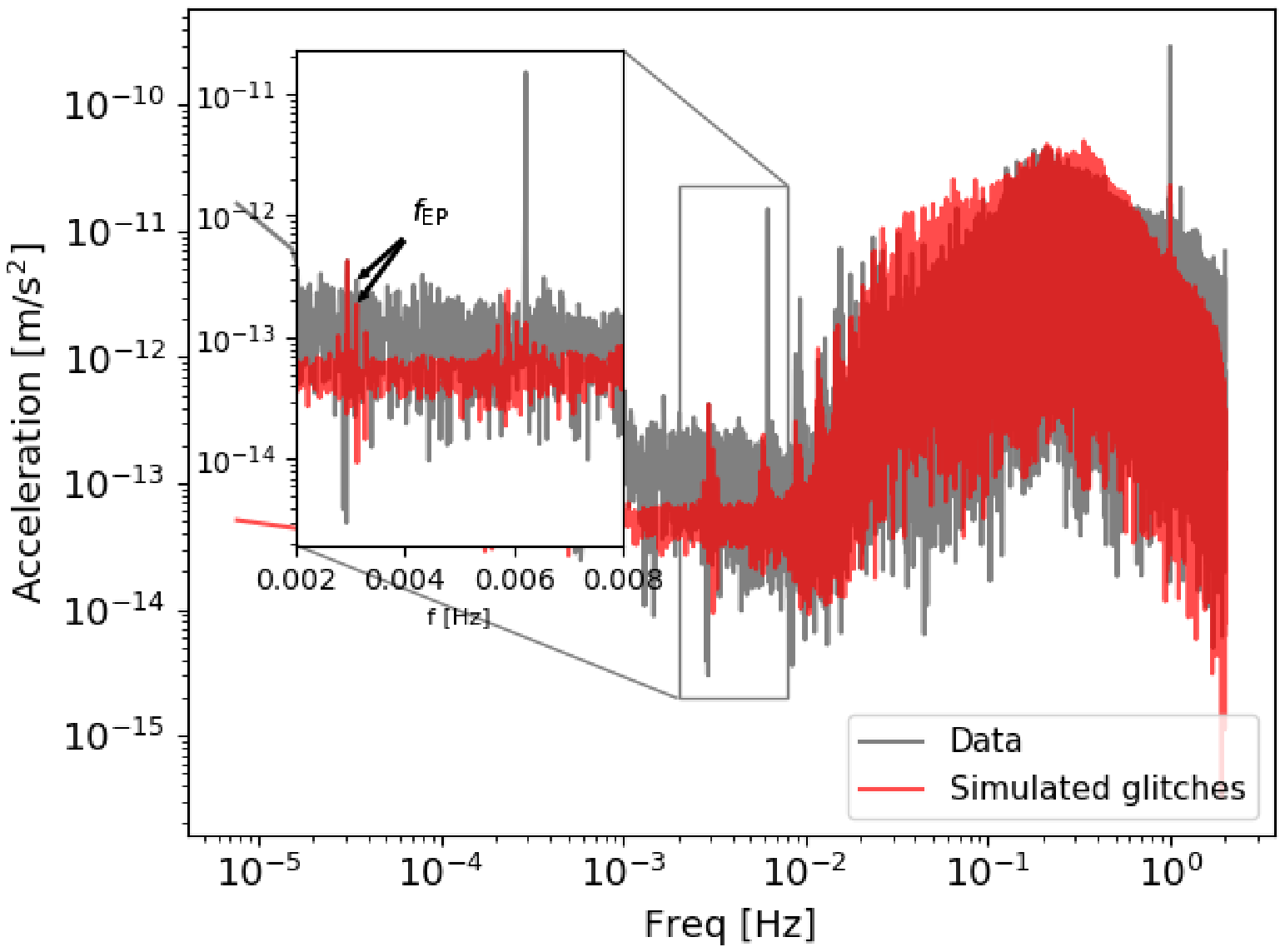}
\includegraphics[width=0.45\textwidth]{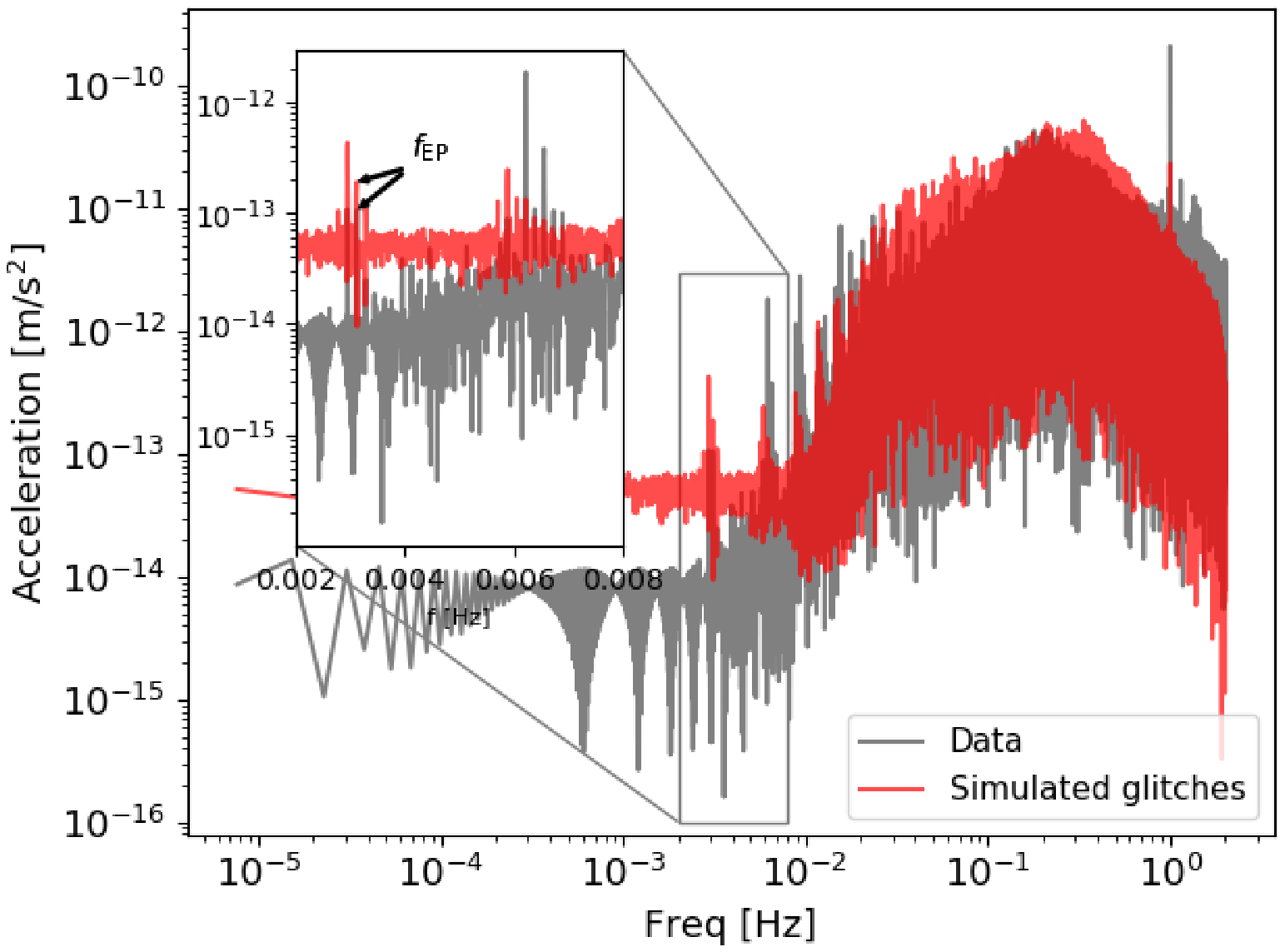}
\caption{Frequency content of glitches computed with our linear model of MICROSCOPE's instrument (red) compared with the measured acceleration (grey); left: out-of-drag-free sensor; right: drag-free sensor). The strong line at $2f_{\rm EP}\approx 0.006$ Hz in the spectrum measured with the out-of-drag-free sensor is the well-known line due to the coupling between the Earth gravity gradient and the offcentering of the sensors, and can be easily and efficiently corrected for \cite{bergecqg7}.}
\label{fig_freqcont}       % Give a unique label
\end{figure}

Fig. \ref{fig_freqcont} shows the frequency content of glitches (Eq. \ref{eq_sg3}) expected from our linear model for both sensors (red), compared with the measured acceleration (grey). The left panel corresponds to the out-of-drag-free sensor (thence, the frequency-dependent low-frequency noise) and the right panel to the drag-free sensor (thence, the flat low-frequency noise); at low frequency, the higher noise level from glitches is due to a conservative assumption about a stochastic background of low SNR glitches. Our model recovers the forests of peaks, though with a level lower than observed, except around the $f_{\rm EP}$ frequency. 

\section{Impact on the test of the WEP} \label{sect_WEP}

\subsection{Linear model} \label{ssect_linearWEP}

Eqs. (\ref{eq_acc1}) and (\ref{eq_Delta}) readily provide an upper bound on the effect of glitches from the acceleration measured at $f_{\rm EP}$ by the sensor at the drag-free point (right panel of Fig. \ref{fig_freqcont}), when assuming a linear model for the instrument.
Indeed, at $f_{\rm EP}$, the acceleration of the drag-free sensor is $M_{\rm df} = E/GA \approx 10^{-13}$m/s$^2$.
Assuming moreover that $dA$ is of order the measured scale factor difference $|dA| \approx 10^{-2}$ \cite{touboul19}, the impact of external forces (including glitches) on the differential acceleration is $\approx 10^{-15}$ m/s$^2$.
In terms of the E\"otv\"os parameter, this amounts to about $1.2\times 10^{-16}$, two orders of magnitude below the 1$\sigma$ uncertainty, and is therefore completely negligible.

\begin{figure}
\center
\includegraphics[width=0.65\textwidth]{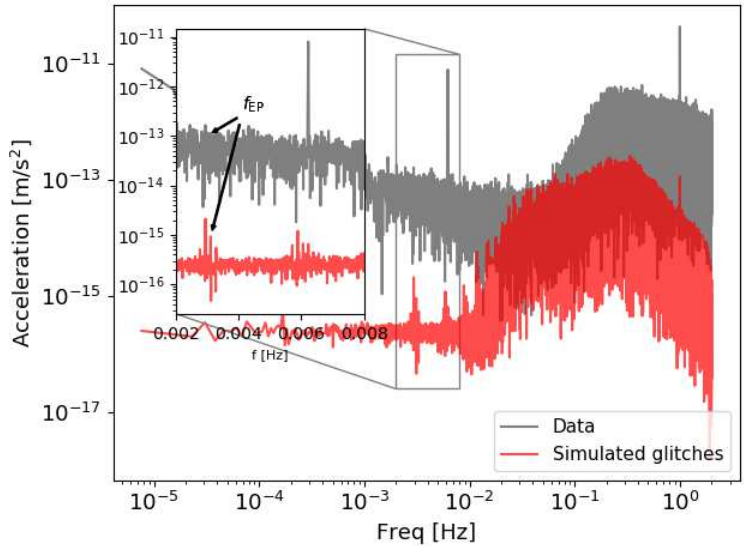}
\caption{Contribution of glitches to the differential acceleration (red) compared with the measured differential acceleration (grey). The strong line at $2f_{\rm EP}\approx 0.006$ Hz in the measured spectrum is the well-known line due to the coupling between the Earth gravity gradient and the offcentering of the sensors, and can be easily and efficiently corrected for \cite{bergecqg7}.}
\label{fig_gcont}       % Give a unique label
\end{figure}

This result can also be seen graphically from the discussion of Sect. \ref{ssect_ffc}. Fig. \ref{fig_gcont} shows the measured differential acceleration (grey) and the differential signal created by glitches, still assuming $|dA| \approx 10^{-2}$. The contribution of glitches at $f_{\rm EP}$, in terms of the E\"otv\"os ratio, can be seen to be at the level of $10^{-16}$, two orders of magnitude below the noise.

\subsection{Confronting reality} \label{ssect_realWEP}

A standard analysis of some measurement sessions provides a statistically unlikely (given the estimates of the E\"otv\"os parameter on other MICROSCOPE measurement sessions \cite{metriscqg9}) level of a WEP violation. One in particular (Session 380 --see Ref. \cite{rodriguescqg4} for a list of sessions) yields a 7$\sigma$ detection on the reference instrument (the test masses of which are made of the same material \cite{liorzou20}). This detection hints towards a systematic effect unaccounted for and made apparent by the lower-than-average noise level of that particular session.

\begin{figure}
\includegraphics[width=0.85\textwidth]{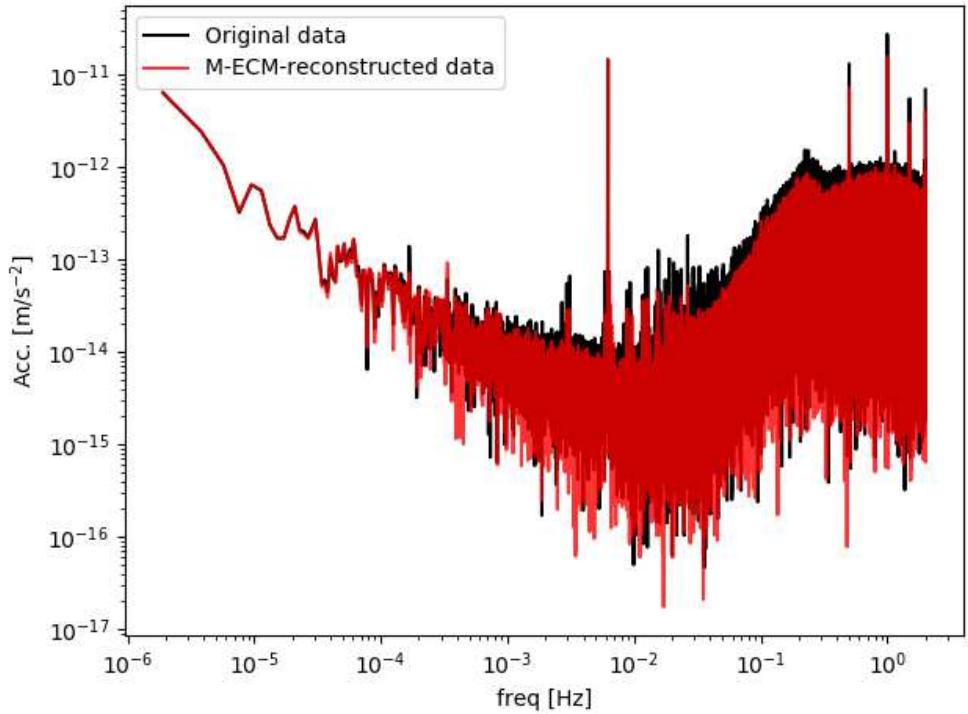}
\caption{Spectra of the $x$ axis differential acceleration for session 380, before (black) and after (red) glitches masking and data reconstruction.}
\label{fig_380}       % Give a unique label
\end{figure}

Despite the fact that our model predicts that glitches have a negligible impact on the WEP test, in light of Sect. \ref{sect_statistics}, it is tempting to attribute this signal to glitches.
To investigate this possibility, we complete our analysis with a heuristic one entirely based on the data.
We mask glitches and reconstruct the masked data with the M-ECM algorithm \cite{baghi16}. M-ECM (Modified-Expectation-Conditional-Maximization) maximises the likelihood of available data through the estimation of missing data by their conditional expectation, based on the circulant approximation of the complete data covariance. For this exercise, we look for glitches not only on the $x$-axis, but also on the $y$- and $z-$axes, to make sure that glitches preferentially projected on one of those latter axes are detected, though buried in the noise of the $x-$axes. Detecting glitches with the $\sigma$-clipping technique described above (using a 4.5$\sigma$ threshold), and masking ten seconds after each outlier, we remove 46\% of the data. This shows that glitches appear in considerable amount and may significantly contribute to the measurement noise. 
Fig. \ref{fig_380} compares the spectra of the raw differential acceleration and of the differential acceleration after discarding glitches and reconstructing the data with M-ECM. The line forest between 0.01Hz and 0.1Hz is significantly reduced, meaning that glitches contribute to it. Some lines are nonetheless still present, hinting to the presence either of lower SNR glitches or of another unidentified contributor.

No significant WEP violation can be detected in the reconstructed data. The aforementioned 7$\sigma$ WEP detection when not masking glitches is therefore directly related to their presence. We checked that the M-ECM-reconstructed data would allow us to accurately estimate a WEP violation in this masked data by adding a mock WEP signal to the original (before masking) data, and correctly estimating it.

\subsection{Discussion}

We developed a linear model of MICROSCOPE's instrument in Sect. \ref{sect_anal}. Feeding it with the distribution of glitches discussed in Sect. \ref{sect_stats}, we qualitatively reproduce the spectral lines in the acceleration measured by each sensor individually, although their amplitudes, except that of the $f_{\rm EP}$ line, are underestimated (Fig. \ref{fig_freqcont}). Relying on this model, we expect that glitches hardly affect the WEP test (Sect. \ref{ssect_linearWEP}). However, this is contradicted by some real-life measurements (Sect. \ref{ssect_realWEP}).

In summary, the model is satisfactory at the level of individual sensors' acceleration (especially about $f_{\rm EP}$), but seems unable to reproduce the signal created by glitches in the differential acceleration. As discussed below, the explanation of this failure can be twofold:
\begin{enumerate}
\item a linear model is too simple and not representative of the complexity of a real system, especially with a measurement system so sensitive that any perturbation is easily spotted
\item the distribution of glitches input to the model is incorrect.
\end{enumerate}

\subsubsection{Non-linear model and internal saturations}

Although they have been estimated to be negligible, quadratic terms $K_{2i}$ appear in MICROSCOPE's measurement equation \cite{rodriguescqg1, bergecqg7}, making the instrument intrinsically weakly non-linear. We added them to our model, but could not find any evidence of a significant effect from a coupling with glitches (lower than $10^{-17}$ at $f_{\rm EP}$).

We managed to mimic significant forests of lines in the power spectrum of the differential acceleration by simulating saturations in the servo-loop electronics which controls the test masses and coupling them with observed glitches, provided that those saturations are asymmetric ({\it i.e.}, they have different characteristics on each test mass). Nevertheless, they should be strong enough to be flagged by the various detectors along the electronics lines, but none was observed. Finally, although they could explain the level of the signal due to glitches in the differential accelerations, it is not clear what their physical origin could be.

\subsubsection{Glitches distribution}

The distribution of glitches passed to our linear model, as discussed in Sect. \ref{sect_stats}, is based on glitches detected above 3$\sigma$ of the noise of a given sensor (we saw that sensors share a globally similar response to glitches, see Fig. \ref{fig_4masses}) on its $x$ axis. However, as discussed in Sect. \ref{ssect_realWEP}, each glitch has a preferential direction and is more or less visible on an axis or another. For instance, glitches detected on the $z$ axis are not always detectable on the $x$ axis and are therefore missed in Sect. \ref{sect_stats}. The distribution of glitches passed to the model in Sect. \ref{ssect_linearWEP} is thus likely incomplete.
As a consequence, the input glitches power may be too small, and the expected level in the differential acceleration power spectrum may be underestimated. Nevertheless, it does not explain why the discrepancy with the observed data should be higher in the differential acceleration than in individual sensors' accelerations. 

Basing the glitches distribution on a 3-axes detection (as done in Sect. \ref{ssect_realWEP}) is also bound to fail. 
First, Fig. \ref{fig_380} shows that in this case, and even when half the data is considered as glitches, the spectral lines supposedly corresponding to glitches are still present (though reduced), hinting toward the presence of many more, very low SNR glitches; it seems unrealistic to detect and take them into account without eventually discarding all the data.
Second, we do not have access to the correct amplitude, on the $x$ axis, of glitches detected on the $y$- and $z$-axes since they are  buried in the noise of the $x$ axis. Yet, this is the very information required to predict the level of signal in the differential acceleration. 
Works similar to that performed for micrometeoroid impacts on LISA Pathfinder \cite{thorpe19} may be required to access this information but they go beyond the scope of this paper and will be planned in the future.

\subsubsection{Conclusion}

Given our current lack of understanding of the exact physical mechanisms at play with glitches (where and how they occur, how they propagate from their origin to the test masses, how they mix with the electronics) and of why our model's performance is poorer with differential accelerations than with individual accelerations, decisively deciphering the weaknesses of our model is still impossible. It is likely that the key lies in a combination of (at least) the two possibilities mentioned above.
For instance, a better glitches detection scheme, {\it e.g.} based on a match-filter technique, may help improve the model of their distribution. Moreover, a simulation of the complete MICROSCOPE satellite system could allow us to better estimate and predict the effect of glitches, and should be the subject of a future work.

Finally, although we cannot compute an {\it a priori} level of the impact of glitches on the WEP test, we definitely showed that glitches have an effect: our best bet is then to mask them out and reconstruct the underlying ``clean" data \cite{metriscqg9}.

\section{Conclusion} \label{sect_ccl}

Glitches in MICROSCOPE data are short-lived events visible in the measured acceleration of MICROSCOPE's test masses. 
In this paper, we investigated their shape and statistics, and their potential impact on the test of the WEP.
They all have the same observed shape (akin to a damped sine) that we can relate to the transfer function of the system made of MICROSCOPE's instrument and drag-free system, meaning that they originate in events that die off quicker than the response time of the measuring apparatus.
However, they are too undersampled and the transfer function not known well enough to allow for an investigation of their ``true" shape.

We also investigated the distribution of their time of arrival and showed that although they are random events, they come with two distinct periods, the satellite's orbital period and its spinning period.
Their cyclic distribution hints toward a link with the Earth albedo, modulated by the sensitivity of each side of the satellite: changes of temperature trigger random crackles in the MLI coating, the occurence of which depends on which side of the satellite faces the Earth. Although this is a reasonable hypothesis, we could not find clear evidence of seasonal effects expected in this case, though more data is necessary to better tackle this task.

Because of those cycles, glitches affect the measured accelerations power spectra, adding a rich frequency content with several forests of spectral lines at well-defined frequencies. In particular, they may create power in the differential acceleration at the WEP test frequency, adding an extra systematic error.
We developed a linear model of the measurement apparatus, taking into account MICROSCOPE's inertial sensors and drag-free system, with the aim to reproduce the frequency content of glitches from the distribution of their time of arrival. We noted that the model reproduces well enough the contribution of glitches in the acceleration measured by each sensor and that glitches should hardly impact the test of the WEP. However, a heuristic analysis shows that reality is more subtle and that the model surely underestimates their contribution to the differential acceleration. This analysis consisted in discarding glitches from a measurement session known to provide a statistically unlikely value of the E\"otv\"os parameter, and filling in the subsequent gaps: we noticed that the estimated E\"otv\"os parameter is strongly reduced when glitches are cut out.

Therefore, while it qualitatively describes the main features of the underlying process, our model does not allow us 
to provide a reliable estimate of the systematic error amplitude.
A much better understanding of the origin and nature of glitches, of their propagation within the satellite's subsystems down to the inertial sensors, and of the instrument itself, may be required to perform this exercise. 
Going beyond a linear model, we found that adding asymmetrical saturations (meaning that both inertia sensors of a differential accelerometer behave differently) allowed us to reproduce the observed differential acceleration spectrum. However, those saturations should happen in parts of the instruments known not to saturate, making their physical meaning and existence hard to explain. Nevertheless, they could hint at a mechanical asymmetry between test masses.
Another source of amplification of the glitches' signal may come from an asymmetry in a sensor's Digital Voltage Amplifier that provides the voltages to be applied to the test mass; investigating this possibility requires in-depth numerical and lab tests, and it will be the subject of a future article.

To sum up, we showed that glitches may create a systematic error in the WEP test, which we cannot robustly predict. However, we also showed that discarding them and filling in the subsequent data gaps mitigates their impact on the E\"otv\"os parameter estimation. Most importantly, we made sure that this masking/filling process does not affect a possible ``real" WEP violation.
Future gravity space missions relying on a technology similar to MICROSCOPE's should make sure that glitches are well constrained, or even totally suppressed by the experimental apparatus. For instance, replacing MICROSCOPE's MLI coating by a rigid coating like GOCE's  should significantly reduce the number of glitches, and therefore their impact on the measurement.

\ack
This work is based on observations made with the T-SAGE instrument, installed on the CNES-ESA-ONERA-CNRS-OCA-DLR-ZARM MICROSCOPE mission. 
We thank Pierre-Yves Guidotti, Jean-Philippe Uzan and Martin Pernot-Borr\`as, as well as the members of the MICROSCOPE Science Working Group for useful discussions and Nicolas Toucquoy for creating preliminary glitches catalogs.
The work of JB, BF, MR and PT is supported by CNES and ONERA fundings. GM acknowledges support from OCA, CNRS and CNES.
JB and SP acknowledge financial support from the UnivEarthS Labex program at Sorbonne Paris Cit\'e (ANR-10-LABX-0023 and ANR-11-IDEX-0005-02), and thank NASA's Goddard Space Flight Center, where part of this work was performed, for hospitality. 
QB is supported by an appointment to the NASA Postdoctoral Program at the Goddard Space Flight Center, administered by Universities Space Research Association under contract with NASA.

\appendix

\section{Glitches vs satellite's geocentric position and attitude} \label{app_geo}

\begin{figure}
\center
\includegraphics[width=0.45\textwidth]{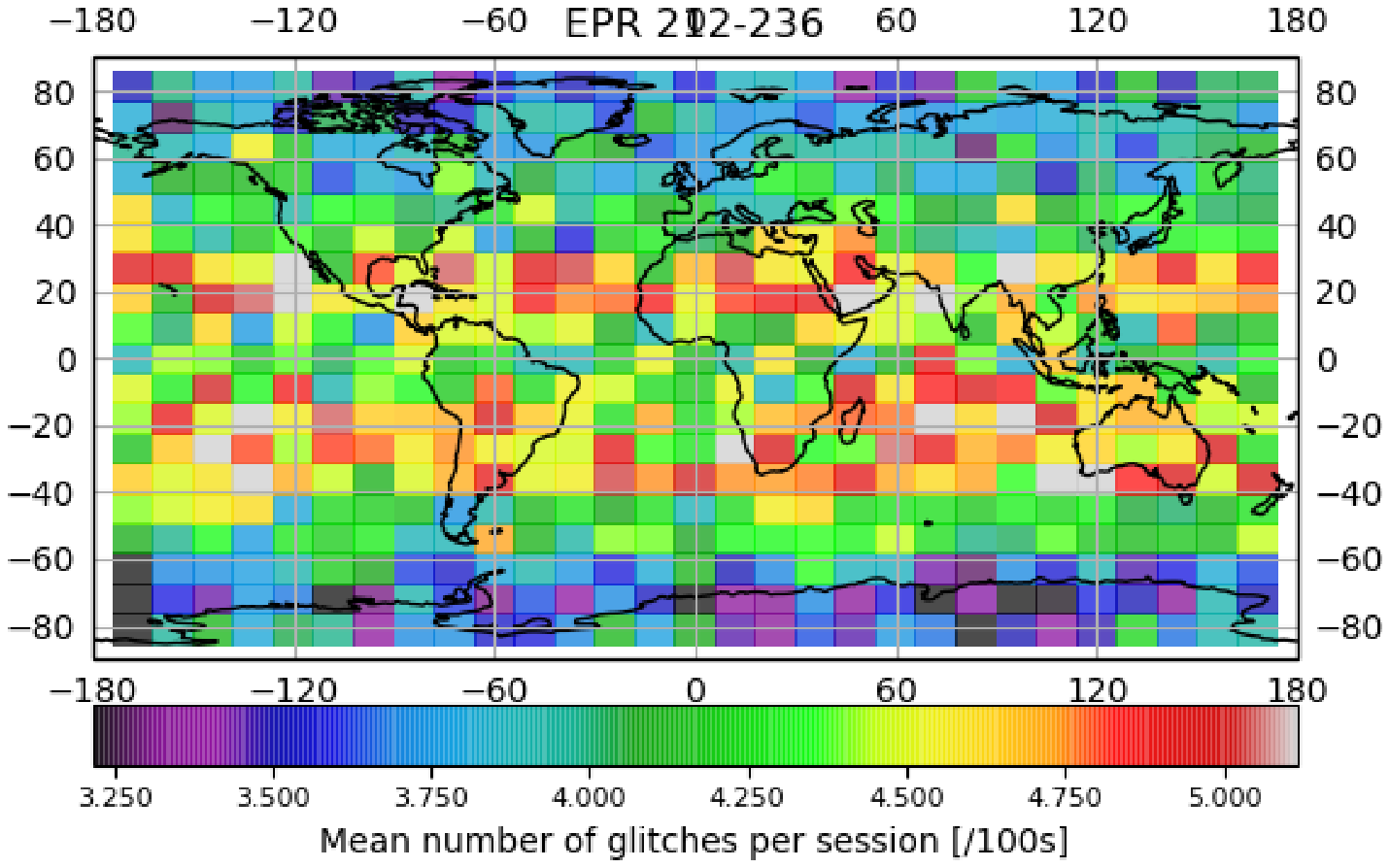}
\includegraphics[width=0.45\textwidth]{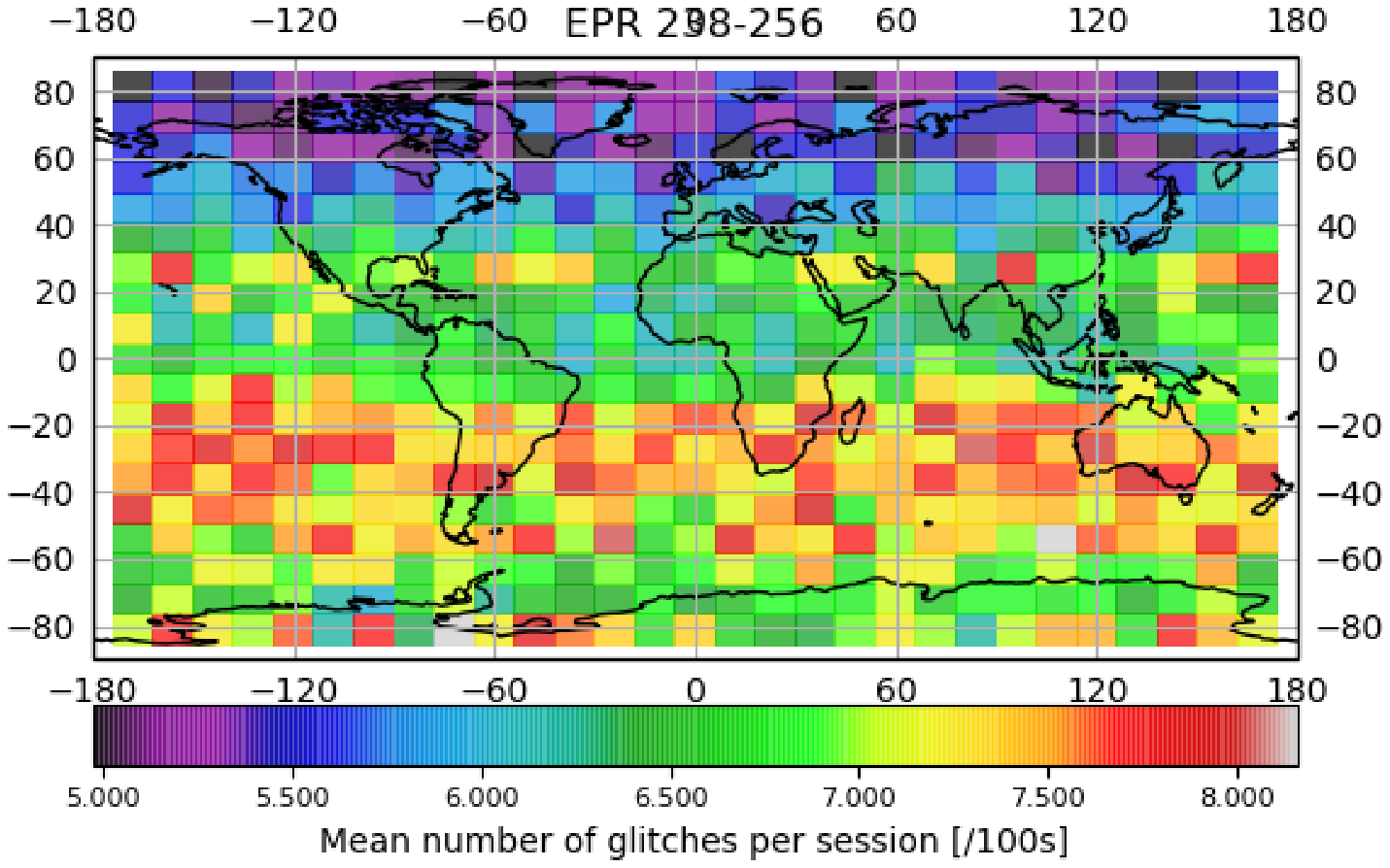}
\caption{Geographical distribution of ${\rm SNR}>3$ glitches (mean number in bins of latitude and longitude), averaged over four measurement sessions at different epochs (January-February 2017 --top-- vs March-April 2017 --bottom).}
\label{fig_geo}       % Give a unique label
\end{figure}

Fig. \ref{fig_geo} shows the geographical distribution of ${\rm SNR}>3$ glitches averaged over four measurements in January-February 2017 (upper panel) and in March-April 2017 (lower panel). The different satellite's spinning characteristics from session to session allow us to marginalise over ``local'' effects (such as the disymmetries between the characteristics of each side of the satellite) and to extract the environmental effects. 
We can notice that the glitches distribution is not homogeneous, but is rather dipolar, with glitches more likely to occur some latitudes, the dipole moving with time.
The geographical distribution of glitches is more homogeneous and centered on the equator in the upper panel of Fig. \ref{fig_geo} than in its lower panel, where it is highly disymmetric with respect to the equator. 

\begin{figure}
\center
\includegraphics[width=0.65\textwidth]{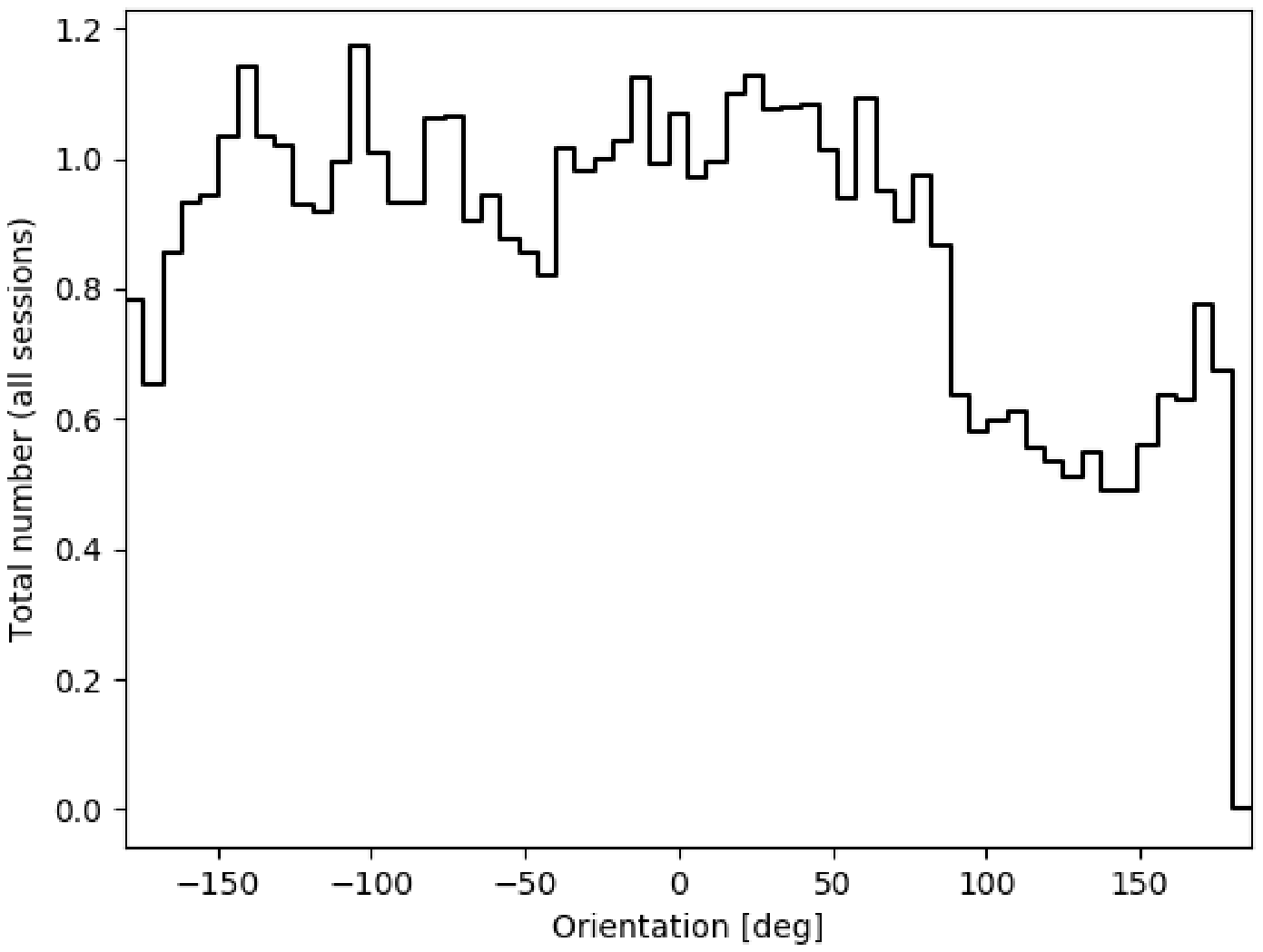}
\caption{Number of glitches as a function of the satellite's attitude.}
\label{fig_orientation}       % Give a unique label
\end{figure}

Fig. \ref{fig_orientation} shows the histogram of glitches as a function of the satellite's attitude, parametrised by the orientation of the $x$-axis of SUEP's test mass with respect to the Earth, averaged over 60 measurement sessions. Clearly, such a histogram marginalises over geographical effects. It is clear that the distribution is not homogeneous, meaning that, in the hypothesis where glitches originate in MLI crackles, all sides of the satellite do not have the same likelihood to crack. This confirms the spin period seen in the 2-point correlation function.

The exact mechanism that trigers glitches is not yet fully understood, but their geographical distribution hints to thermal effects from the Earth albedo and the Sun illumination, though the Solar weather may also contribute (this will be investigated in a future work), with those thermal effects affecting the MLI coating of each side of the satellite in a different manner.

\section*{References}
\bibliographystyle{iopart-num}
\bibliography{mic18}

\end{document}